\documentclass[USenglish, onecolumn]{scrartcl}
\usepackage[USenglish]{babel}
\usepackage[a4paper, margin=2cm, bottom=3cm]{geometry}

\usepackage[utf8]{inputenc}
\usepackage{lmodern}
\usepackage{microtype}

\usepackage{authblk}

\usepackage{abstract}
\newcommand{\placeabstract}[1]{\begin{abstract}\noindent{}#1\end{abstract}\twocolumn}

\newcommand{\qt}[1]{`#1'}

\newcommand{\dir}{./}
\newcommand{\ldir}{\dir literature.bib}
\newcommand{\gdir}{\dir Graphics/}

\usepackage[
    backend=biber,
    style=numeric-comp,
  	maxalphanames=1,
    sortlocale=en_US,
    natbib=true,
    url=false, 
    doi=false,
    eprint=true,
    giveninits=true,
  	maxbibnames=5,
    maxcitenames=1,
    sorting=none
]{biblatex}

\DefineBibliographyStrings{USenglish}{
  andothers = {et al.}
}
\renewbibmacro*{volume+number+eid}{
  \printfield{volume}\iffieldundef{volume}{}{\mkbibbold{\addcomma}}
  \thefield{number}
}
\renewbibmacro*{journal+issuetitle}{%
  \usebibmacro{journal}%
  \setunit*{\addspace}%
  \iffieldundef{series}
    {}
    {\newunit
     \printfield{series}%
     \setunit{\addspace}}%
  \usebibmacro{volume+number+eid}%
  \newunit%
}
\DeclareBibliographyDriver{article}{%
  \usebibmacro{bibindex}%
  \usebibmacro{begentry}%
  \usebibmacro{author/translator+others}%
  \setunit{\labelnamepunct}\newblock
  \usebibmacro{title}%
  \newunit
  \printlist{language}%
  \newunit\newblock
  \usebibmacro{byauthor}%
  \newunit\newblock
  \usebibmacro{bytranslator+others}%
  \newunit\newblock
  \printfield{version}%
  \newunit\newblock
  \usebibmacro{in:}%
  \usebibmacro{journal+issuetitle}%
  \newunit
  \usebibmacro{byeditor+others}%
  \newunit
  \usebibmacro{note+pages}%
  \setunit{\addspace}%
  \usebibmacro{issue+date}%
  \setunit{\addcolon\space}%
  \usebibmacro{issue}%
  \newunit\newblock
  \usebibmacro{doi+eprint+url}%
  \newunit\newblock
  \usebibmacro{addendum+pubstate}%
  \setunit{\bibpagerefpunct}\newblock
  \usebibmacro{pageref}%
  \usebibmacro{finentry}}
\renewbibmacro{in:}{%
  \ifentrytype{article}{}{\printtext{\bibstring{in}\intitlepunct}}%
}

\AtEveryBibitem{\clearfield{month}}
\AtEveryBibitem{\clearfield{day}}
\AtEveryBibitem{%
  \ifentrytype{misc}{%
  }{%
    \clearfield{eprint}%
  }%
}
\DeclareNameAlias{author}{last-first}
\DeclareNameAlias{editor}{last-first}
\DeclareNameAlias{translator}{last-first}
\DeclareFieldFormat[article]{title}{{#1}}
\DeclareFieldFormat[inproceedings]{title}{{#1}}
\DeclareFieldFormat[misc]{title}{{#1}}
\DeclareFieldFormat[article]{volume}{\mkbibbold{#1}}
\DeclareFieldFormat[article]{number}{\mkbibparens{#1}}
\DeclareFieldFormat{pages}{#1}
\usepackage{soulutf8}

\usepackage{amsmath}
\usepackage{amssymb}
\usepackage{bm}
\usepackage{nicefrac}
\usepackage{mathtools}
\usepackage{siunitx}
\usepackage{algorithm}
\usepackage{algpseudocode}

\usepackage{hyperref}

\newboolean{OnlyCaptions}
\setboolean{OnlyCaptions}{false}
\newboolean{NoCaptions}
\setboolean{NoCaptions}{false}

\usepackage{graphicx}
\ifthenelse{\not\boolean{NoCaptions}}{%
	\usepackage[labelfont=bf, format=plain]{caption}%
	\captionsetup[figure]{name={Fig.}}%
	\captionsetup[table]{name={Tab.}}%
}{%
	\renewcommand{\caption}[1]{}%
}
\addto\extrasUSenglish{%
}
\newcommand{\subfignum}[1]{#1}
\newcommand{\subfiglabel}[1]{\textbf{\subfignum{#1}}}
\newcommand{\subfigref}[2]{\autoref{#1}\subfignum{#2}}

\NewDocumentCommand{\fig}{s O{1} O{} O{} O{} O{figure} m}{%
	\ifthenelse{\not\boolean{OnlyCaptions}}{%
		\begin{\IfBooleanTF #1{#6*}{#6}}[#4]%
			\centering%
			\includegraphics[width = #2\textwidth]{\gdir #7}%
			\caption[#5]{#3}%
			\label{fig:#7}%
		\end{\IfBooleanTF #1{#6*}{#6}}%
	}{#3}%
}

\begin{document}	

\hyphenation{neg-a-tive-ly-charged wave-guide Borregaard nano-beam}

\title{`Sawfish' Photonic Crystal Cavity for Near-Unity Emitter-to-Fiber Interfacing in Quantum Network Applications}

\author[1,2]{Julian~M.~Bopp}
\author[3]{Matthias~Plock}
\author[1]{Tim~Turan}
\author[1]{Gregor~Pieplow}
\author[3,4]{Sven~Burger}
\author[1,2,*]{Tim~Schröder} 
\affil[1]{Humboldt-Universität zu Berlin, Department of Physics, 12489 Berlin, Germany}
\affil[2]{Ferdinand-Braun-Institut gGmbH, Leibniz-Institut für Höchstfrequenztechnik, 12489 Berlin, Germany}
\affil[3]{Zuse Institute Berlin (ZIB), 14195 Berlin, Germany}
\affil[4]{JCMwave GmbH, 14050 Berlin, Germany}
\affil[*]{Corresponding author: Tim Schröder, tim.schroeder@physik.hu-berlin.de}
\maketitle

\placeabstract{%
Photon loss is one of the key challenges to overcome in complex photonic quantum applications.
Photon collection efficiencies directly impact the amount of resources required for measurement-based quantum computation and communication networks.
Promising resources include solid-state quantum light sources, however, efficiently coupling light from a single quantum emitter to a guided mode remains demanding.
In this work, we eliminate photon losses by maximizing coupling efficiencies in an emitter-to-fiber interface.
We develop a waveguide-integrated `Sawfish' photonic crystal cavity and use finite element simulations to demonstrate that our system transfers, with $\num{97.4}\,\%$ efficiency, the zero-phonon line emission of a negatively-charged tin vacancy center in diamond adiabatically to a single-mode fiber.
A surrogate model trained by machine learning provides quantitative estimates of sensitivities to fabrication tolerances.
Our corrugation-based design proves robust under state-of-the-art nanofabrication parameters, maintaining an emitter-to-fiber coupling efficiency of $\num{88.6}\,\%$.
To demonstrate its potential in reducing resource requirements, we apply the Sawfish cavity to a recent one-way quantum repeater protocol.
}
	
\section{Introduction}
Photon losses play a critical role in realizing quantum technology applications \citep{Chen2021, Neuman2021}. In quantum communication protocols \citep{Borregaard2020} and optical one-way quantum computation \citep{Raussendorf2001}, the losses must not exceed a particular threshold. Losing photons in long-distance quantum communication applications directly limits information transfer rates as well as maximally bridgeable communication distances to approximately $\SI{100}{\kilo\meter}$ \citep{Krutyanskiy2019}.
As opposed to classical communication, the distance cannot be extended compensating losses by signal amplification due to the no-cloning theorem \citep{Wootters1982}.
Quantum repeaters, however, overcome the no-cloning theorem by exploiting entanglement \citep{Briegel1998}. One-way quantum repeaters use multi-photon encoding and quantum error-correcting codes to protect quantum information from losses and other errors \citep{Fowler2010, Munro2012, Muralidharan2014, Azuma2015, Glaudell2016, Ewert2016, Lee2019}.

The amount of photons $N_\mathrm{ph}$ required for such multi-photon entangled states strongly depends on the employed encoding protocol and the efficiency of transferring photons from one network node to adjacent nodes. Recently, Borregaard \textit{et al.} \citep{Borregaard2020} proposed an efficient one-way quantum repeater protocol that requires node-to-node (emitter-to-detection) efficiencies of around $\num{95}\,\%$ depending on the system error rate $\epsilon_\mathrm{r}$ \citep{Borregaard2020}. The emitter-to-detection efficiency includes the photon generation and fiber coupling on the sender and the photon detection efficiency on the receiver site.
In turn, the photon generation involves an interface between a stationary atom-like system and photons acting as `flying qubits'. A photonic crystal cavity generally lies at the heart of this interface. It enhances light-matter interaction by means of the Purcell effect \citep{Purcell1946} and thus the emitter's emission rate \citep{Rugar2021, Kuruma2021}.
Likewise, the cavity improves photon collection efficiencies \citep{Knall2022}.
Although significant effort has been spent towards controlling and increasing light-matter interaction \citep{Chen2021, Knall2022}, it remains an open question how to design interfaces that provide the collection efficiencies required by next generation quantum repeater protocols, including the aforementioned proposal by J. Borregaard \textit{et al.} \citep{Borregaard2020}.

\begin{figure*}[h]
	\centering
	\includegraphics[scale=1]{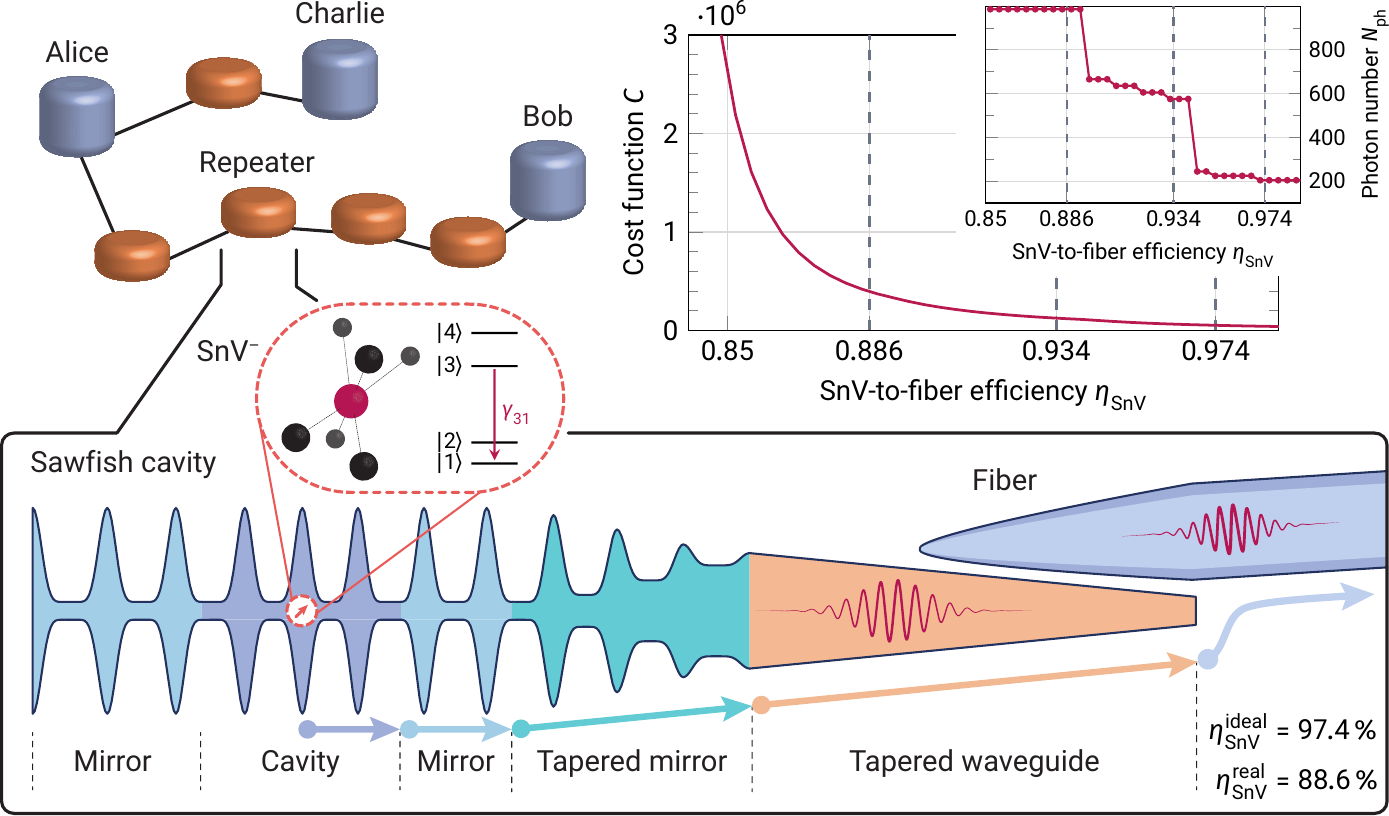}
	\caption{	
	\textbf{Overview of the fiber-coupled Sawfish cavity as a building block for quantum networks.}
	Several participants (blue nodes) communicate over a quantum network. Repeater stations (orange nodes) bridge long distances between the participants.
	Each repeater station is assumed to be based on a quantum repeater according to a scheme by J. Borregaard \textit{et al.} \citep{Borregaard2020} with SnV$^-$ centers embedded in Sawfish cavities as sources for quantum communication resource states.
	The SnV$^-$ is enclosed by two mirror regions in the optical cavity's center enabling Purcell enhancement of the $\gamma_{31}$ transition.
	In the sketch, the left mirror possesses a higher reflectivity than the right mirror. The latter transitions into a tapered region to firstly couple part of the emitted light adiabatically to a waveguide and subsequently to a fiber.
	The inset displays the repeater scheme's cost function $C$ and the tree-cluster resource state's required photon number $N_\mathrm{ph}$ as functions of the SnV-to-fiber efficiency (equation \eqref{eq:interface_efficiency}) for a reencoding error $\epsilon_\mathrm{r}=\num{0.1}\,\text{\textperthousand{}}$. Vertical dashed lines at $\eta_\mathrm{SnV}^\mathrm{real} = \num{88.6}\,\%$ correspond to state-of-the-art cavity quality factors $2Q\approx\num{17000}$ \citep{Mouradian2017} (see \autoref{ss:exp_soa_conditions}), lines at $\eta_\mathrm{SnV} = \num{93.4}\,\%$ to $2Q\approx\num{76000}$, which is expected to be within reach in the close future, and lines at $\eta_\mathrm{SnV}^\mathrm{real} = \num{97.4}\,\%$ to an idealized cavity design.
	}
	\label{fig:App}
\end{figure*}

In this work, we propose a nanophotonic cavity-to-fiber interface based on a novel `Sawfish' design (\autoref{fig:App}) which reveals near-unity emitter-to-fiber coupling efficiencies. We benchmark its performance evaluating the cost function defined by Borregaard \textit{et al.} \citep{Borregaard2020}
\begin{equation}
	C = \frac{1}{\Gamma_\mathrm{tcs} f p_\mathrm{trans}(\eta)}\frac{m L_\mathrm{att}}{ \tau_\mathrm{ph} L} \,,
	\label{eq:costBorr}
\end{equation}
as a function of the emitter-to-fiber collection efficiency $\eta$. $\Gamma_\mathrm{tcs}$ is the tree-cluster state generation rate, $f$ the secret-bit fraction of the transmitted qubits \citep{Scarani_2009}, $p_\mathrm{trans}(\eta)$ the transmission probability that depends on $\eta$ and on the tree-cluster state configuration, $m$ the number of repeater stations, $L_\mathrm{att}$ the optical fiber attenuation length, $\tau_\mathrm{ph}$ the photon emission time, and $L$ the total communication distance. $1/C$ can be interpreted as the secret key rate in units of the photonic qubit emission time per repeater station and per attenuation length for a given total distance $L$ \citep{Borregaard2020}. The detection efficiency is implicitly included in $p_\mathrm{trans}(\eta)$ (see supplementary information for details).  

With the Purcell factor $F_\mathrm{P}$, we define the emitter-to-fiber coupling efficiency $\eta$ as
\begin{align}
	\eta = \beta_\mathrm{C}\!\left(F_\mathrm{P}\right)\times\beta_\mathrm{WG}\times\beta_\mathrm{F}\times DW\!\left(F_\mathrm{P}\right)\,,
	\label{eq:interface_efficiency}
\end{align}
where $\beta_\mathrm{C}\!\left(F_\mathrm{P}\right)=F_\mathrm{P}/\left(F_\mathrm{P}+1\right)$ is the probability to emit a photon into the cavity mode \citep{Reiserer2015, Tomm2021}, $\beta_\mathrm{WG}$ the cavity-to-waveguide coupling efficiency, and $\beta_\mathrm{F}$ the waveguide-to-fiber coupling efficiency. By multiplication with the emitter-specific and Purcell factor-dependent \citep{Li2015} Debye-Waller factor $DW\!\left(F_\mathrm{P}\right)$, we restrict the emitter-to-fiber efficiency to the ratio of photons emitted into the respective emitter's zero-phonon line (ZPL) \citep{Gaebel2004}. Hence, $\eta$ describes the overall probability that a single photon is emitted into a solid-state emitter's ZPL and then successfully transferred to an optical fiber. $\eta$ drastically affects the cost function $C$ and therefore the cluster state size $N_\mathrm{ph}$ that is best suited for distributing a secret key between two distant parties (insets in \autoref{fig:App}).
While $C$ is smoothly decreasing, $N_\mathrm{ph}$ decreases stepwise with increasing system efficiency underlining that photon losses have to be reduced to a particular threshold. 
The cost reduction with increasing emitter-to-fiber efficiency becomes intuitive considering that lower losses result in fewer photons needed for error correction. This leads to higher state generation rates $\Gamma_\mathrm{tcs}$ as well as potentially fewer repeater stations.

For the strong dependence $C\!\left(\eta\right)$, we optimize the Sawfish cavity and its emitter-to-fiber interface to examine the repeater performance under ideal and experimentally feasible efficiencies. As a case study, we consider the negatively-charged tin vacancy center in diamond (SnV$^-$). The SnV$^-$ possesses a ZPL at $\SI{484.3}{\tera\hertz}$ and a comparably large ground state splitting of $\SI{850}{\giga\hertz}$ \citep{Iwasaki2017} rendering it a promising optically-active spin emitter \citep{Goerlitz2022}. Our spin-photon interface, however, is applicable to a large variety of solid-state quantum emitters allowing for its integration with nanophotonic devices based on different material platforms.

The Sawfish cavity design provides several advantages compared to present designs. Particularly, it enables for the first time near-unity emitter-to-fiber coupling efficiencies of up to $\num{97.4}\,\%$. In contrast to the multitude of existing photonic crystal cavity geometries which rely on tiny hole-like features `drilled' into a (suspended) waveguide \citep{Mouradian2017, Evans2018, Rugar2021, Kuruma2021}, the Sawfish cavity is based on large-size corrugation features and a low-loss adiabatic taper region. For gentle mode confinement \citep{Akahane2003} leading to high cavity quality factors $Q$ as well as for adiabatic cavity-to-waveguide coupling \citep{Burek2017, Knall2022}, the cavity features have to be precisely adjusted.
For the latter, this is required to convert Bloch modes in the taper region into waveguide modes.
Avoiding mode mismatch-induced scattering losses at cavity-waveguide interfaces is not straight forward since it necessitates arbitrary control of the cavity feature sizes \citep{Palamaru2001, Sauvan2005}.
Thus, \qt{conventional} hole-based geometries suffer from fabrication difficulties in tapering off their hole diameters adiabatically \citep{Palamaru2001}. In turn, hole-based Bloch waveguides do not provide arbitrarily smooth transitions into unperturbed waveguide regions. Our corrugated design overcomes this limit as it relies on open and relatively large features that can be tapered down adiabatically.
We show that tailoring the photonic potential barrier formed by the Sawfish cavity's taper region allows for waveguide-coupled cavities with distinct cooperativities without reducing their cavity-to-waveguide coupling efficiencies. For instance, tuning of the cooperativity and the cavity-waveguide coupling is needed to maximize the state fidelity within the PEPSI scheme used to map photonic polarization-encoded qubits onto solid-state spins \citep{Chen2021}.
Furthermore, replacing holes with corrugation features minimizes undesired surface effects by increasing the distance between an embedded emitter and the cavity's surface. As a consequence, the Sawfish cavity provides fabrication uncertainty-tolerant near-unity collection of light generated by e.g. a diamond color center through an adiabatic cavity-to-waveguide followed by an adiabatic waveguide-to-fiber transition.
Even for cavities with realistic and thus relatively high loss rates, the emitter-to-fiber coupling efficiency hardly decreases.

After optimizing the performance of the emitter-to-fiber coupling, we explore how nanofabrication tolerances affect the Sawfish cavity. We develop and apply a surrogate modeling procedure which yields information on how specific fabrication parameter distributions influence the expected performance. Vice versa, the procedure reveals parameter distributions minimally required to make a performance parameter, like the Purcell factor, exceed a certain threshold. To showcase this modeling technique, we discuss the effect of non-ideal emitter placement and deviations in the cavity's thickness, width, and side wall angles. Besides gauging the robustness of the Sawfish design in presence of deviations from its ideal configuration, surrogate modeling aids to anticipate which aspects of the device fabrication require the highest precision. To achieve comparability with experimentally realized cavity designs, we consider realistic cavity parameters. In particular, we employ the highest experimentally achieved diamond photonic crystal cavity quality factor $Q$ to date reported on \citep{Mouradian2017} to provide a fair benchmark and to outline how close state-of-the-art emitter-to-fiber interfaces approach the stringent requirements of one-way quantum repeaters.

Finally, we study the efficacy of the Sawfish cavity loaded with a variety of color centers such as the negatively-charged nitrogen (NV$^-$) \citep{Doherty2013} or negatively-charged group-IV \citep{Bradac2019} vacancy centers in diamond.
Within the frame of the quantum repeater protocol by J. Borregaard \textit{et al.}, we demonstrate how perspective improvements in nanofabrication methods give rise to significantly enhanced cluster state generation rates.

\section{Results}
Firstly, we outline the physical considerations, which inform the design of the Sawfish cavity. Secondly, a tapered mirror region is added to the cavity to adiabatically convert the confined cavity mode into a traveling waveguide mode. This mode is next transferred to a tapered fiber. Finally, we investigate the cavity design's robustness under experimental state-of-the-art conditions such as non-optimal emitter placement and fabrication tolerances.
 
\subsection{Cavity design}
\begin{figure*}
	\centering
	\includegraphics[scale=.9]{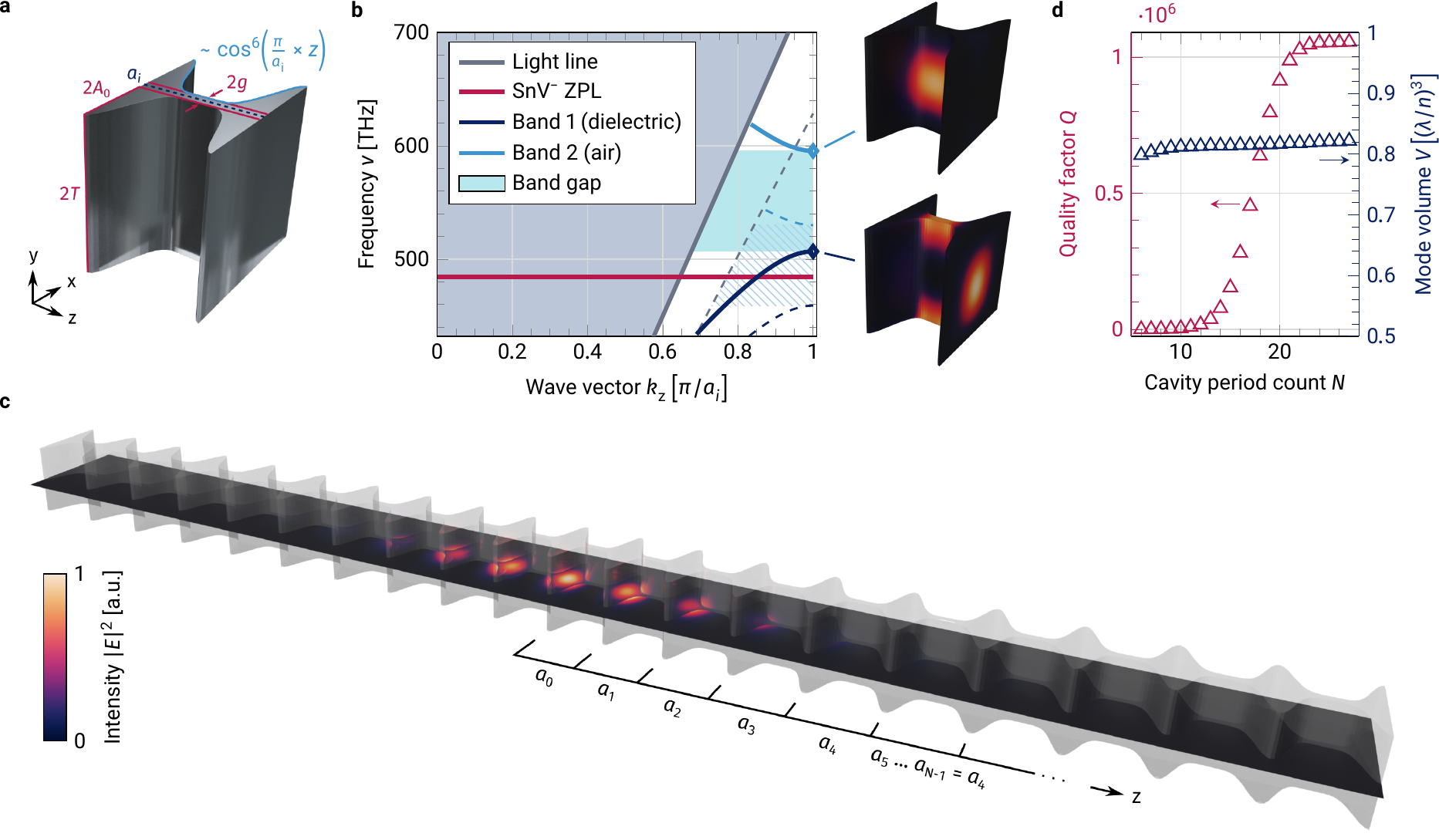}
	\caption{\textbf{Sawfish unit cell and cavity design.}
	\subfiglabel{a} Unit cell of the Sawfish cavity with its characteristic parameters. $a_i$ describes its lattice constant, $2T$ its thickness, $2 A_0$ the (untapered) amplitude of a $\cos^6$ function offset by $g$ from the $x=0$ symmetry plane (dashed line) defining the unit cell's profile.
	\subfiglabel{b} Band diagram of a periodic array of Sawfish unit cells showing TM-like modes. Light modes inside the light cone (gray-shaded area given by $\nu\geq c\left| k\right|$ with $c$ being the speed of light) may escape the structure. Other components propagate as Bloch modes indicated by blue solid and dashed lines for lattice constants $a_0 = \SI{200}{\nano\meter}$ and $a_4 = \SI{238.5}{\nano\meter}$, respectively. Between the fundamental Bloch mode (dark blue line) and the next higher air mode (light blue line), there is a $\SI{88.7}{\tera\hertz}$-wide band gap (blue-shaded area) for a unit cell lattice constant of $a_0$. For this lattice constant, the SnV$^-$ ZPL emission (red line) is guided as the fundamental mode. Emission is suppressed for a unit cell lattice constant of $a_4$ by the ZPL falling into the band gap (blue-dashed area). The insets portray the electric field intensity of the modes highlighted with blue diamonds.
	\subfiglabel{c} Sawfish cavity geometry. The two symmetric halves of the cavity consist of $N$ concatenated Sawfish unit cells each with lattice constants increasing from $a_0$ to $a_4$. The center plane displays the electric field intensity.
	\subfiglabel{d} Convergence behavior of the cavity's quality factor $Q$ (red) and its mode volume $V$ (blue) depending on the amount of concatenated unit cells counting from on the cavity's central $z=0$ symmetry plane.}
	\label{fig:Cavity}
\end{figure*}

For constructing a cavity based on corrugation features, a periodic structure formed by the cavity's unit cell must result in a photonic band gap. The band gap has to be large enough to reliably suppress the transmission of light around its center frequency. We perform eigenmode simulations to determine the Sawfish unit cell's dispersion properties and to identify band gaps centered around the SnV$^-$ ZPL. For diamond with a refractive index of $n=\num{2.41}$ at $\SI{484.3}{\tera\hertz}$ \citep{Phillip1964}, we find that a unit cell with lattice constant $a_i$ and a sinusoidal corrugation along the $z$ direction proportional to $\cos^e\!\left(\pi/a_i\times z\right)$ with $e=\num{2}$ (parameters as indicated in \subfigref{fig:Cavity}{a}) yields a gap-midgap ratio of only $\num{7.8}\,\%$. Raising the exponent $e$ increases the gap-midgap ratio dramatically due to a higher refractive index contrast between the unit cell's edges and its center. For $e=\num{6}$ with all other parameters unchanged, a gap-midgap ratio of $\num{16.1}\,\%$ is obtained. We choose $e=\num{6}$ since larger exponents do not affect the band gap significantly.
A larger exponent would instead reduce the distance between an emitter located at the center of one of the unit cell's rectangular interfaces and the unit cell's curved surface. A small emitter-surface distance potentially promotes undesired interactions between the emitter's dipole and surface charges \citep{Chakravarthi2021a, OrphalKobin2022}.

With the parameters $T=\SI{133}{\nano\meter}$, $A_0=\SI{65}{\nano\meter}$, $g=\SI{11}{\nano\meter}$, blue solid and dashed lines in \subfigref{fig:Cavity}{b} show the photonic band diagram for a periodic structure built by concatenating unit cells with lattice constants $a_0=\SI{200}{\nano\meter}$ and $a_4=\SI{238.5}{\nano\meter}$, respectively. The unit cells are optimized to maximize the band gap, to permit the transmission of the SnV$^-$ ZPL for the lattice constant $a_0$, and to prohibit the transmission for $a_4$. Unit cells with lattice constants $a_0$ guide the ZPL emission as a dielectric mode which is mostly confined in the diamond material. Larger lattice constants shift the band diagram towards lower frequencies eventually making the ZPL emission approach the band gap's center. Unit cells with lattice constant $a_4$ hence act as mirror elements. 
The cavity's center region gently confines a light mode in the vicinity of the SnV$^-$ ZPL if $N$ unit cells with increasing lattice constants $a_0, a_1, \dots a_{N-1}$ are arranged on either side around the cavity's $z=0$ symmetry plane. We reach the highest quality factor $Q=\num{1.06e6}$ with $\left[ a_1,\allowbreak{} a_2,\allowbreak{} a_3\right] = \left[ \SI{207.0}{\nano\meter},\allowbreak{} \SI{217.9}{\nano\meter},\allowbreak{} \SI{228.4}{\nano\meter}\right]$ in addition to $a_0$ and $a_4$ as mentioned above. $a_5$ to $a_{N-1}$ equal $a_4$. \subfigref{fig:Cavity}{c} visualizes the resulting electric field intensity in the cavity.

With the cavity's quality factor $Q$ and its mode volume $V\!$, the overall emission rate of an emitter embedded into the cavity is increased by the Purcell factor $F_\mathrm{P}\sim Q/V$ compared to its bulk emission rate \citep{Purcell1946}. A high quality factor in combination with a small mode volume therefore also raises the ZPL emission through an increased Debye-Waller factor $DW(F_\mathrm{P})$ and hence the emitter-to-fiber coupling efficiency $\eta$.
Since each mirror element constituting the cavity contributes to a photonic potential barrier \citep{Lin1994}, the quality factor rapidly increases for a critical amount of mirror elements as soon as tunneling of the cavity mode through the barrier approaches full suppression (\subfigref{fig:Cavity}{d}). In our case, simulation-based state-of-the-art \citep{Bayn2014, Mouradian2017} quality factors above $\num{1e6}$ can be reached for $N > 20$. The mode volume stays comparably low at $\num{0.8}\,\left(\lambda/n\right)^3$ over the full range of investigated cavity period counts with $\lambda$ being the individual wavelength of the respective cavity's resonance mode.

As an example, we have chosen the cavity parameters to accommodate the SnV$^-$ ZPL at $\SI{484.3}{\tera\hertz}$. However, the design concept outlined here can easily be adapted to other defect centers as well as to completely different material systems. Larger gap widths $g$ might increase the stability of fabricated devices in trade for slightly narrower band gaps.

\subsection{Cavity to waveguide coupling}
\begin{figure*}
	\centering
	\includegraphics[scale=1]{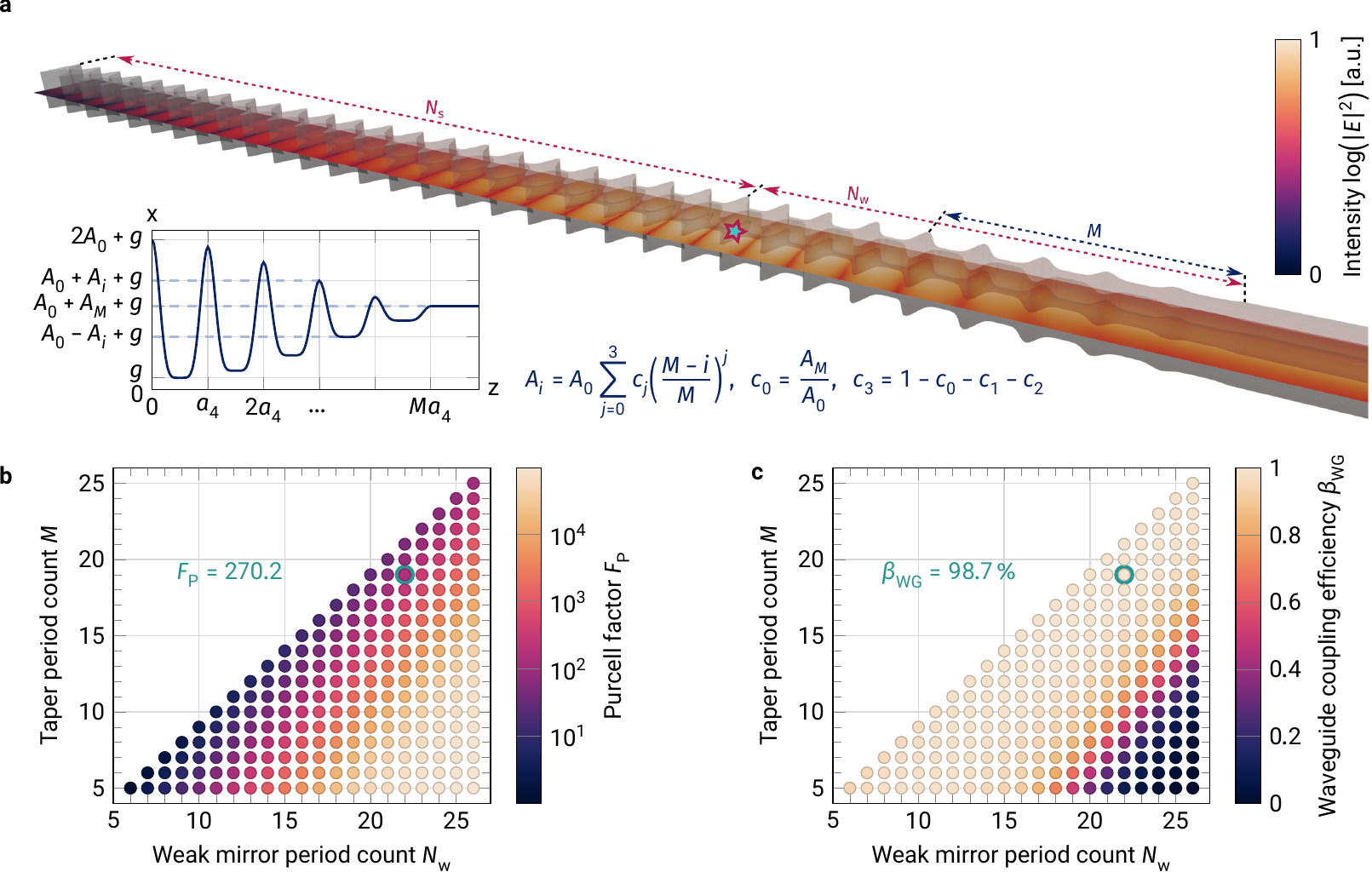}
	\caption{\textbf{Waveguide-coupled Sawfish cavity design.}
	\subfiglabel{a} The waveguide-coupled cavity consists of a strong mirror region of $N_\mathrm{s}$ unit cells with amplitude $2A_0$ left of the emitter indicated by the small star and a weak mirror region of $N_\mathrm{w}$ unit cells right of the emitter. The amplitudes $2A_i$ ($i\in\left[0,M\right]$) of the unit cells within the weak mirror region stay constantly at $2A_0$ for $N_\mathrm{w}-M$ periods (beginning at the emitter) and taper down to the final amplitude $A_0+A_M$ according to a third-order polynomial with coefficients $c_j$ along the taper region $M$ as depicted in the inset. The center plane displays the logarithm of the electric field intensity.
	\subfiglabel{b} Purcell factor $F_\mathrm{P}$ and \subfiglabel{c} waveguide coupling efficiency $\beta_\mathrm{WG}$ depending on the amount of unit cells forming the weak mirror and on the length of the taper region. $N_\mathrm{s}$ is fixed to $\num{23}$. The data point highlighted in green denotes the configuration with the highest waveguide coupling efficiency. For each point, the [010]-oriented dipole's emission frequency has been set to the respective configuration's resonance.
	}
	\label{fig:WG}
\end{figure*}

Controlled out-coupling of the light mode confined in the cavity requires to weaken the cavity's reflectivity at one side of the emitter. In addition to lowering the reflectivity, we damp the amplitude of the corrugation features to adiabatically transfer the Bloch mode to a waveguide mode. Hence, an asymmetric cavity consisting of a weak and a strong mirror needs to be constructed. The weak mirror consists of $N_\mathrm{w}$ unit cells, the strong mirror of $N_\mathrm{s}$ cells, respectively. The unit cells forming the weak mirror stay at a constant amplitude for $N_\mathrm{w}-M$ periods until they taper down to a waveguide with total width $2\left(A_0+A_M+g\right)$ over $M$ periods according to the profile given by a third-order polynomial as stated in \subfigref{fig:WG}{a}. $A_M$ determines the waveguide's width. Barely affecting the cavity-to-waveguide coupling efficiency, it can be chosen to fit the subsequent fiber coupling requirements. We find by Bayesian optimization \citep{movckus1975bayesian} that a profile following a third-order-polynomial with coefficients $c_1=\num{0.275}$, $c_2=\num{2.243}$ and $A_M=\SI{2.5}{\nano\meter}$ is most efficient for our Sawfish geometry.

For now, we focus on an ideal system with $N_\mathrm{s}=\num{23}$ and a dipole perfectly overlapping with the desired TM-like mode. A corresponding untapered symmetric cavity with $N=23$ possesses a central frequency $\omega_\mathrm{c}=2\pi\times\SI{484.70}{\tera\hertz}$ and an intrinsic scattering loss rate $\kappa_\mathrm{s} = \omega_\mathrm{c}/Q_\mathrm{s} = 2\pi\times\SI{463.8}{\mega\hertz}$ (\autoref{fig:Cavity}). By choosing weak mirrors with different amounts of contributing unit cells $N_\mathrm{w}$ and different taper lengths $M$\!, the Purcell factor $F_\mathrm{P}$ and thus the emitter-cavity cooperativity can be adjusted over a wide range (\subfigref{fig:WG}{b}). The waveguide coupling efficiency $\beta_\mathrm{WG}$ defined as the ratio of the Poynting vector integrated over the waveguide's end facet to the Poynting vector integrated over the entire computational domain remains at approximately $\beta_\mathrm{WG}=\num{99}\,\%$ over the parameter range significant for cooperativity tuning (\subfigref{fig:WG}{c}). In turn, the Purcell factor might be arbitrarily chosen without causing additional scattering losses at the tapered region.

For each weak mirror period count $N_\mathrm{w}$, there is a range of taper lengths which maximize both, the Purcell factor as well as the waveguide coupling efficiency. In contrast, the Purcell factor drops if $M$ is chosen too high for the cavity-to-waveguide loss rate becoming excessive. If $M$ is chosen too low, $F_\mathrm{P}$ increases, but suddenly $\beta_\mathrm{WG}$ drops for the cavity-to-waveguide loss rate being reduced to the intrinsic loss rate $\kappa_\mathrm{s}$. Such dependencies emphasize the need to thoroughly design tapered regions interfacing cavities with waveguides and not to neglect their influence on the cavities themselves.

To experimentally investigate group-IV defects in diamond like the SnV$^-$, often highly pure CVD diamond samples are ion-implanted and subsequently annealed \citep{Iwasaki2017, Rugar2021, Kuruma2021}. The respective off-the-shelf diamond substrates are usually cut along the crystallographic $\left[100\right]$ direction causing dipoles to enclose an angle of $\num{54.7}^\circ$ with each cartesian axis ($\left[111\right]$ direction) \citep{Hepp2014, Goerlitz2020}. Consequently, for modeling a realistic scenario, the effect of rotating the dipole embedded in our Sawfish design within a plane spanned by the $\left[111\right]$ ($\alpha=\num{54.7}^\circ$) and e.g. the $\left[010\right]$ ($\alpha=\num{0}^\circ$) direction has to be examined. We observe that the Purcell factor follows a $F_\mathrm{P}\sim g^2\sim\cos^2\!\left(\alpha\right)$ law since the coupling constant $g$ depends on a scalar product between the cavity mode at the dipole's position and its orientation. A realistic dipole orientation reduces the Purcell factor to approximately $\num{33}\,\%$ of its value obtained for an ideal orientation. Remarkably, the waveguide coupling efficiency is hardly affected. It is slightly reduced to $\num{99.8}\,\%$ of its ideal value (see supplementary information).

\subsection{Waveguide to fiber coupling}
\begin{figure*}
	\centering
	\includegraphics[scale=1]{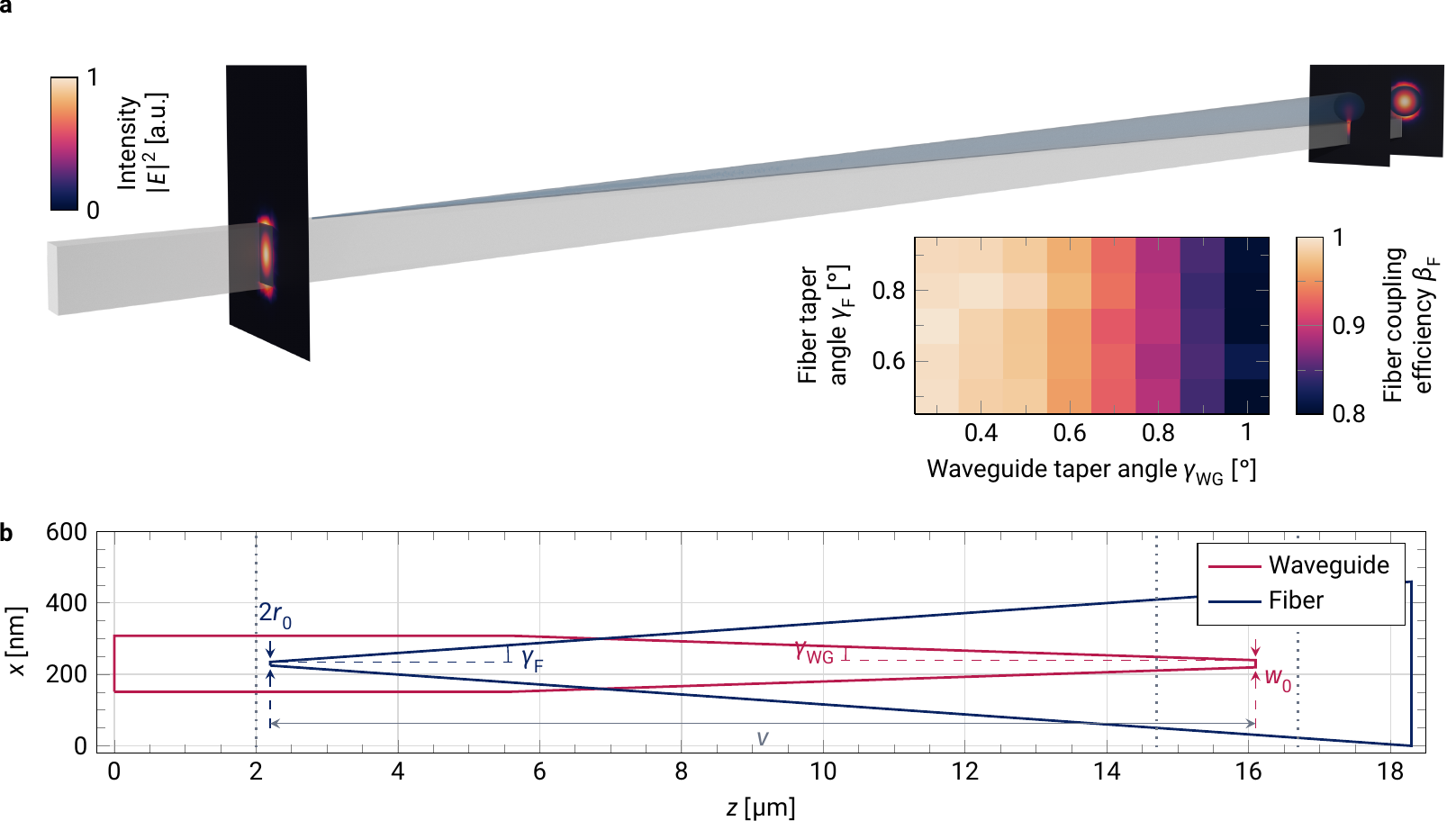}
	\caption{\textbf{Fiber-coupling of the waveguide attached to the Sawfish cavity.}
	\subfiglabel{a} Simulated model of the tapered waveguide (gray-shaded) coupled to a tapered fiber placed on top (blue-shaded) including three cross-sectional views of the electric field intensity. The inset shows the dependence of the waveguide-to-fiber transmission $\beta_\mathrm{F}$ on the taper angles as defined in \subfiglabel{b}. The length of the overlapping region $v$ has been optimized for each set of angles.
	\subfiglabel{b} Top view of the waveguide-fiber system. A diamond waveguide (red) with a thickness of $2T$ and an initial total width of $2\left(A_0 + A_M + g\right)$ as it is attached to the Sawfish cavity is tapered down under an angle $\gamma_\mathrm{WG}$ to a final width $w_0$. A tapered fiber (blue) modeled as a conical frustum with a final radius $r_0$ and a taper angle $\gamma_\mathrm{F}$ overlaps the tapered waveguide for a length $v$ measured from the waveguide's tip. Gray dotted lines denote $z$ coordinates of the planes displaying cross-sectional intensities in \subfiglabel{a}.
	}
	\label{fig:Fiber}
\end{figure*}

Aiming to eventually couple light generated by a cavity-coupled emitter into a fiber network, we show that the waveguide attached to our Sawfish cavity allows for efficient fiber coupling. We use a waveguide-fiber interface according to \citep{Tiecke2015}. The waveguide narrows down to its final width $w_0=\SI{5}{\nano\meter}$ under an angle $\gamma_\mathrm{WG}$. A conical fiber tip with a core refractive index of $n=\num{1.463}$ tapered down to a radius $r_0=\SI{5}{\nano\meter}$ under an angle $\gamma_\mathrm{F}$ is put on top of the waveguide overlapping it for a length $v$ (\autoref{fig:Fiber}).

A waveguide-to-fiber transmission $\beta_\mathrm{F}$ of up to $\beta_\mathrm{F}=\num{99.4}\,\%$ is reached for $\gamma_\mathrm{WG}=\num{0.3}^\circ$, $\gamma_\mathrm{F}=\num{0.7}^\circ$ and an overlap $v=\SI{20}{\micro\meter}$ (inset in \subfigref{fig:Fiber}{a}). The effective refractive index of the optical mode traveling through the combined system initially resembles the effective refractive index of the fundamental waveguide mode, but approaches the one of the fundamental fiber mode emphasizing successful evanescent coupling \citep{Tiecke2015} (see supplementary information).

\subsection{Dipole displacement and fabrication tolerances}
\label{ss:fab_tolerances}
Until this point, our investigation of the Sawfish cavity was mainly based on ideal conditions disregarding scattering losses due to e.g. fabrication tolerances or rough surfaces. To explore how fabrication tolerances impact the Sawfish cavity's performance, we perform Monte Carlo sampling on a surrogate model. Once the surrogate model is trained with computationally expensive simulation data for a set of fabrication parameters, any uncertainty distribution of the selected parameters can be evaluated quickly (see supplementary information). Lacking the need for further computationally expensive simulations when exchanging the uncertainty distributions renders the surrogate modeling procedure a powerful tool to efficiently analyze the impact of fabrication tolerances on any nanostructure. From now on, we consider realistically $\left[111\right]$-oriented dipoles.

\begin{figure*}
	\centering
	\includegraphics[scale=.9]{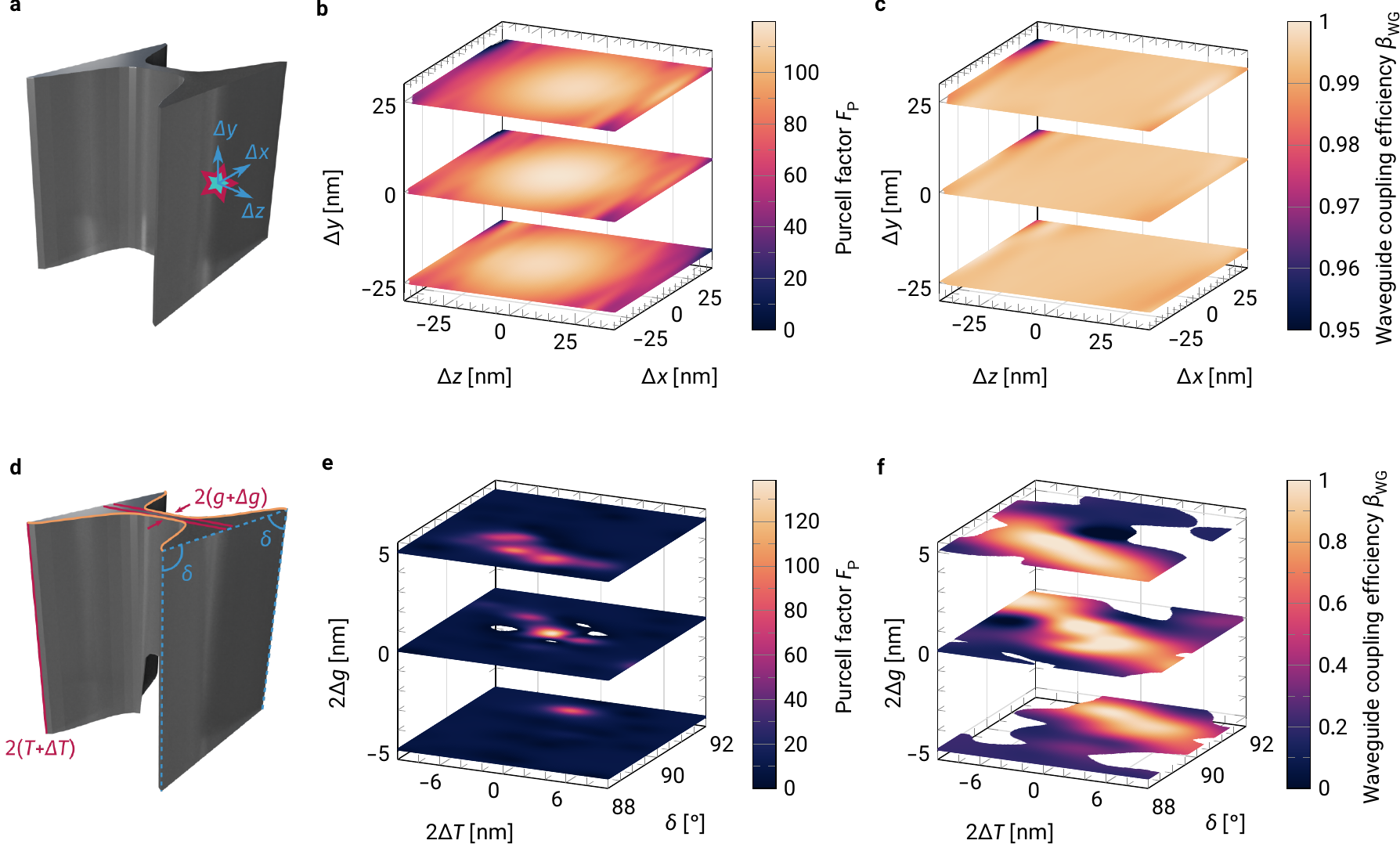}
	\caption{\textbf{Influence of dipole displacement and fabrication tolerances.}
	\subfiglabel{a} For ideal fabrication parameters, the $\left[111\right]$-oriented dipole (indicated as a star) is displaced along the cartesian axes by distances $\Delta x$, $\Delta y$, and $\Delta z$ measured from the cavity's center point.
	\subfiglabel{b}, \subfiglabel{c} Colors describe the resulting Purcell factor $F_\mathrm{P}$ (\subfiglabel{b}) and waveguide coupling efficiency $\beta_\mathrm{WG}$ (\subfiglabel{c}) for displaced dipoles.
	\subfiglabel{d} With an ideally positioned, $\left[111\right]$-oriented dipole, fabrication geometry parameters are altered. The cavity's thickness is varied by $2\Delta T$ and its gap width by $2\Delta g$. Its side wall angle $\delta$ is allowed to deviate from $\num{90}^\circ$. Orange lines denote edges for which the influence of rounded corners is investigated.
	\subfiglabel{e}, \subfiglabel{f} Colors describe the Purcell factors $F_\mathrm{P}$ (\subfiglabel{e}) and waveguide coupling efficiencies $\beta_\mathrm{WG}$ (\subfiglabel{f}) for deviations from ideal fabrication geometry parameters. Predictions by the surrogate model with high uncertainties have been removed from the displayed planes.
	}
	\label{fig:Uncertainties}
\end{figure*}

With a dipole displaced by distances $\left[\Delta x,\allowbreak{} \Delta y,\allowbreak{} \Delta z\right]$ scattering around a zero mean value (ideal dipole position) with standard deviations of $\SI{25}{\nano\meter}$ each (\subfigref{fig:Uncertainties}{a}), we obtain $\overline{F}_{\mathrm{P}} = \left( \num{91.4} \pm \num{3.2} \right)^{+\num{18.3}}_{-\num{26.0}}$ and $\overline{\beta}_{\mathrm{WG}} = \left( \num{0.9863} \pm \num{0.0014} \right)^{+\num{0.0003}}_{-\num{0.0014}}$ as median values for the Purcell factor and the waveguide coupling efficiency, respectively. The values' uncertainties denote combined Monte Carlo and surrogate uncertainties. Lower and upper bounds indicate lower and upper standard deviations of the values' distributions. \subfigref{fig:Uncertainties}{b} and \subfignum{c} present the dependencies $F_\mathrm{P}\left(\Delta x,\Delta y, \Delta z\right)$ and $\beta_\mathrm{WG}\left(\Delta x,\Delta y, \Delta z\right)$ as predicted by the surrogate model.
While the Purcell factor only diminishes for high deviations along the $x$- and $z$-axis, it is even less affected by dipoles displaced along the $y$-axis. This effect is understandable considering the cavity mode profile (inset in \subfigref{fig:Cavity}{b}). The electric field spreads further in $y$-direction compared to the other directions causing the mode overlap to be less affected by dipole displacements in this direction. In contrast, $\beta_\mathrm{WG}$ is hardly influenced by any displacement.

Likewise, we investigate fabrication geometry parameters $\left[\Delta T,\allowbreak{} \delta,\allowbreak{} \Delta g \right]$ distributed around their ideal design values $\left[\SI{0}{\nano\meter},\allowbreak{} \SI{90}{\degree},\allowbreak{} \SI{0}{\nano\meter} \right]$ (\subfigref{fig:Uncertainties}{d}). Rounded corners along the edges highlighted with orange lines in \subfigref{fig:Uncertainties}{d} as another relevant parameter do not have any influence up to a corner rounding radius of at least $\SI{5}{\nano\meter}$. \subfigref{fig:Uncertainties}{e} and \subfignum{f} display predictions of the single fabrication parameters' impact on $F_\mathrm{P}$ and $\beta_\mathrm{WG}$, respectively. In general, $\beta_\mathrm{WG}$ is affected much less than $F_\mathrm{P}$ by fabrication tolerances. However, the Purcell factor strongly depends on the correct choice of fabrication parameters. This is the case since fabrication uncertainties shift the unit cells' band structures away from the cavity condition rapidly. The system then becomes a pure waveguide supporting Bloch modes, but no resonant cavity mode, which is expressed in regions with low Purcell factors but still high waveguide coupling efficiencies. Some parameters reveal correlations. For instance, a cavity with a lowered side wall angle works efficiently if the gap is enlarged at the same time. Thus, we emphasize the need for precise and controllable nanofabrication techniques.

For standard deviations $\left[\SI{0.8}{\nano\meter},\allowbreak{} \SI{0.1}{\degree},\allowbreak{} \SI{0.8}{\nano\meter} \right]$ of the fabrication geometry parameters $\left[\Delta T,\allowbreak{} \delta,\allowbreak{} \Delta g \right]$, we obtain $\overline{F}_{\mathrm{P}} = \left( \num{48.5} \pm \num{26.5} \right)^{+\num{48.0}}_{-\num{33.5}}$ and $\overline{\beta}_{\mathrm{WG}} = \left( \num{0.929} \pm \num{0.054} \right)^{+\num{0.057}}_{-\num{0.141}}$.  $\overline{F}_{\mathrm{P}}$ and $\overline{\beta}_{\mathrm{WG}}$ describe median values for the ratio of $\num{95}\,\%$ working Sawfish cavities (see supplementary information) assuming the given fabrication tolerances and cavity tuneability within a frequency range between $\SI{483.6}{\tera\hertz}$ and $\SI{485.1}{\tera\hertz}$.

\subsection{Experimental state-of-the-art conditions}
\label{ss:exp_soa_conditions}
In the previous section, we assessed how fabrication tolerances explicitly arising from displaced dipoles or deviating fabrication parameters impact the Sawfish cavity. In this section, we instead investigate the influence of reduced quality factors without defining the origin of the respective additional loss channel. Combining both approaches allows to relate desired quality factors and thus also Purcell factors to minimally required dipole displacement distributions and fabrication tolerances.

\begin{figure*}
	\centering
	\includegraphics[scale=1]{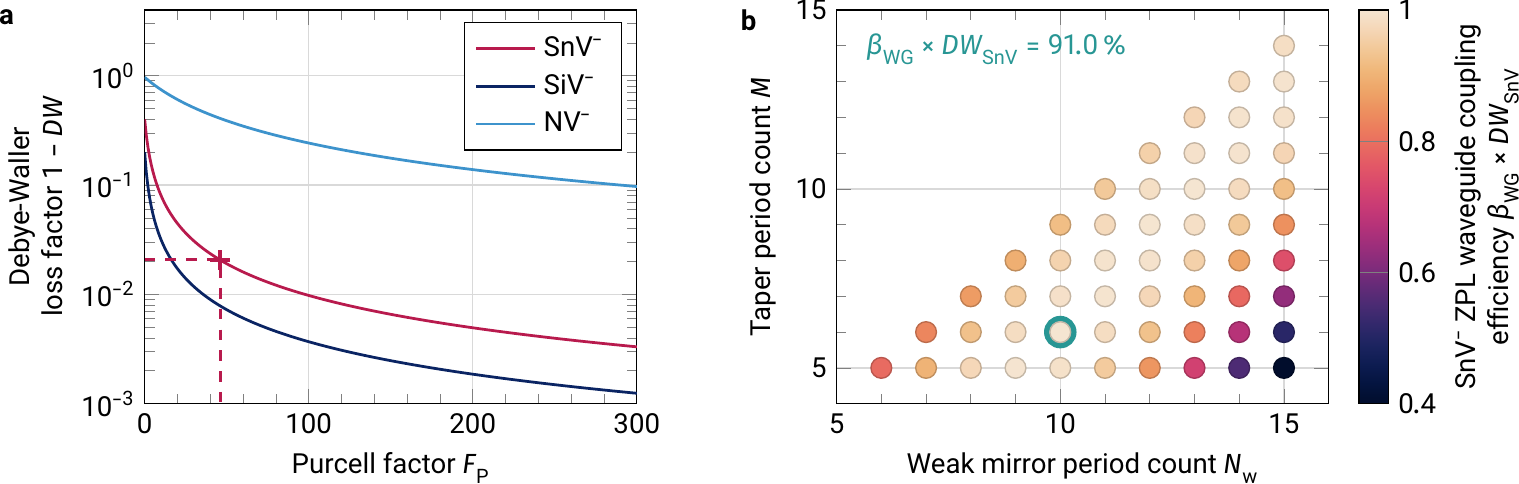}
	\caption{\textbf{Debye-Waller factor and ZPL waveguide coupling efficiency for a $\bm{\left[111\right]}$-oriented SnV$^-$ and realistic losses.}
	\subfiglabel{a} Dependence of different diamond defect centers' Debye-Waller factors $DW$ on the Purcell factor $F_\mathrm{P}$. The red cross at $F_\mathrm{P}=\num{46.4}$ highlights the SnV$^-$-based Sawfish cavity configuration with the highest efficiency for realistic losses (green circle in \subfiglabel{b}). The GeV$^-$'s Debye-Waller factor is not depicted since it matches the SnV$^-$'s curve.
	\subfiglabel{b} SnV$^-$ ZPL waveguide coupling efficiency $\beta_\mathrm{WG}\times DW_\mathrm{SnV}$ for different amounts of unit cells forming the weak mirror and different lengths of the taper region. The encircled data point denotes the configuration with the highest efficiency. The dipole has been adjusted to the cavity's resonance for each point.
	}
	\label{fig:DW}
\end{figure*}

To represent experimental state-of-the-art conditions without modeling every possible fabrication uncertainty explicitly, we fix the quality factor of an untapered symmetric Sawfish cavity to $2Q\approx\num{17000}$ by reducing its length to $N_\mathrm{s}=N_\mathrm{w}=\num{11}$ (\subfigref{fig:Cavity}{d}). This value is in the order of magnitude of a currently fabricated photonic crystal cavity \citep{Mouradian2017}.
The factor $\num{2}$ takes into account that half of the light leaving a symmetric cavity will be collected through the weak mirror of a corresponding tapered asymmetric cavity. The collected part does not contribute to the losses. It is coupled to the attached waveguide instead.
Next, we introduce such a tapered region to couple the cavity with intrinsic scattering losses and a realistic $\left[111\right]$-oriented dipole to a waveguide. For the best compatible taper region as highlighted in \autoref{fig:DW}, we obtain $F_\mathrm{P}=\num{46.4}$ and $\beta_\mathrm{WG}=\num{92.9}\,\%$ which agree well with the fabrication geometry parameters $\left[\Delta T,\allowbreak{} \delta,\allowbreak{} \Delta g \right]$ distributed as described in \autoref{ss:fab_tolerances}. Hence, surrogate modeling allows to directly map a desired quality factor to requirements on fabrication tolerances.

The Purcell factor directly determines $\beta_\mathrm{C}$ as well as the Debye-Waller factor $DW_\mathrm{SnV}$. With the emission rate into the ZPL $\gamma_{31}$ (inset in \autoref{fig:App}), the emission rate into the phonon sideband (PSB) $\gamma_\mathrm{PSB}$ and the SnV$^-$'s natural Debye-Waller factor in bulk $DW_\mathrm{SnV}^0\approx\num{60}\,\%$ \citep{Goerlitz2020}, its Debye-Waller factor in presence of Purcell enhancement becomes
\begin{equation}
	DW_\mathrm{SnV}\left(F_\mathrm{P}\right) = \frac{\gamma_{31}F_\mathrm{P}\left(1-DW_\mathrm{SnV}^0\right) + DW_\mathrm{SnV}^0 \gamma_\mathrm{PSB}}{\gamma_{31}F_\mathrm{P}\left(1-DW_\mathrm{SnV}^0\right) + \gamma_\mathrm{PSB}}
\end{equation}
as plotted in \subfigref{fig:DW}{a} (see supplementary information). With an optimized taper region (\subfigref{fig:DW}{b}), we estimate the overall emitter-to-fiber collection efficiency of a lossy realistic system for the SnV$^-$ ZPL to $\eta_\mathrm{SnV}^\mathrm{real} = \num{88.6}\,\%$ neglecting further systemic optical losses \citep{Knall2022}. For ideal conditions (configuration highlighted in \subfigref{fig:WG}{b} and \subfignum{c}), the collection efficiency rises to $\eta_\mathrm{SnV}^\mathrm{ideal} = \num{97.4}\,\%$. Within the frame of the quantum repeater protocol by J. Borregaard \textit{et al.} \citep{Borregaard2020}, such increases in $\eta_\mathrm{SnV}$ reduce the cluster state size $N_\mathrm{ph}$ and raise the cluster state generation rate $\Gamma_{\mathrm{tcs}}$ by an order of magnitude as shown in the inset of \autoref{fig:App} and as detailed later in \autoref{fig:Repeater}.

\section{Discussion}
\begin{figure}[h]
	\centering
	\includegraphics[scale=1]{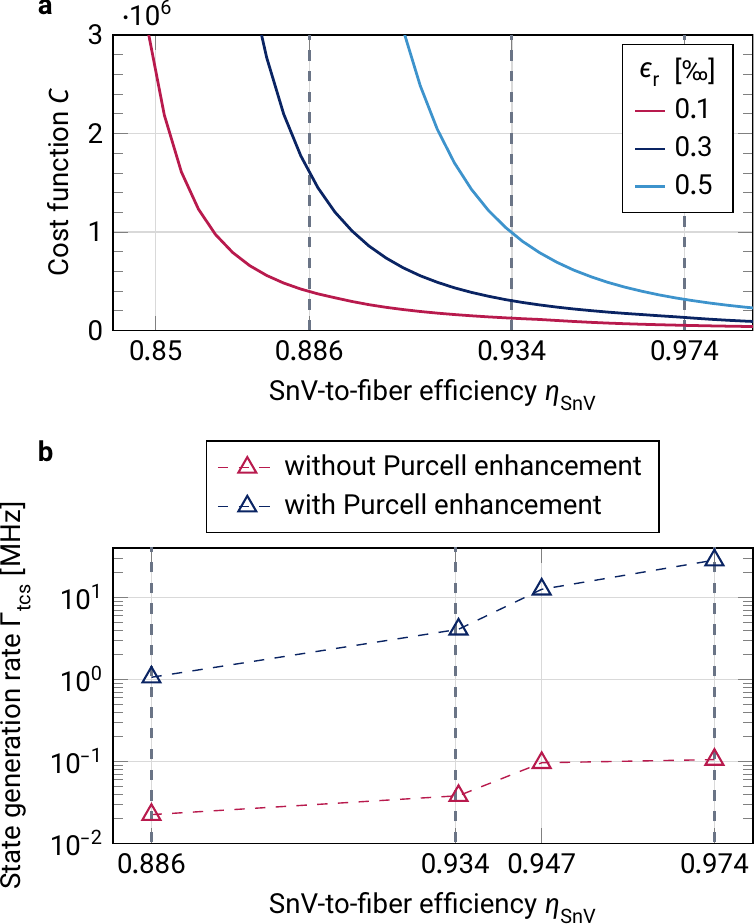}
	\caption{\textbf{Efficiency of the quantum repeater scheme \citep{Borregaard2020} based on a SnV$^-$-loaded Sawfish cavity.}
	\subfiglabel{a} Cost function $C$ (equation \eqref{eq:costBorr}) and \subfiglabel{b} tree cluster state generation rate $\Gamma_{\mathrm{tcs}}$ (see supplementary information) as functions of the SnV-to-fiber efficiency $\eta_\mathrm{SnV}$.
	The cost function is shown for different reencoding errors $\epsilon_\mathrm{r}$ while the reencoding error is fixed to $\epsilon_\mathsf{r}=\num{0.1}\,\text{\textperthousand{}}$ for the tree cluster state generation rate.
	Red and blue dashed lines serve as guides to the eye. Vertical dashed lines match the corresponding dashed lines in the insets of \autoref{fig:App} denoting cavities with different intrinsic scattering losses.
	}
	\label{fig:Repeater}
\end{figure}

We introduced and optimized a waveguide-coupled `Sawfish cavity' and estimated the overall probability of successfully transferring ZPL photons of an embedded SnV$^-$ to an optical fiber. For ideal conditions, a probability of $\eta_\mathrm{SnV}^\mathrm{ideal}=\num{97.4}\,\%$ was obtained which dropped for state-of-the-art cavity losses \citep{Mouradian2017} only slightly to $\eta_\mathrm{SnV}^\mathrm{real}=\num{88.6}\,\%$. Applying a Monte Carlo sampling method performed on a surrogate model enabled us to map specific cavity losses to a set of corresponding fabrication tolerances. Even taking these scattering losses into account, the overall efficiency $\eta_\mathrm{SnV}^\mathrm{real}$ exceeds to date findings in recent work for hole-based photonic crystal cavities \citep{Knall2022}. We highlighted the benefits of extremely efficient spin-photon interfaces by applying the Sawfish cavity to the one-way quantum repeater scheme of J. Borregaard \textit{et al.} \citep{Borregaard2020}. The repeater nodes' resource requirements, like the tree-cluster state sizes, reduced significantly (inset in \autoref{fig:App}).
Comparing the repeater's cost function (equation \eqref{eq:costBorr}) for different reencoding errors (\subfigref{fig:Repeater}{a}) supports these findings.
In every case, higher overall emitter-to-fiber efficiencies drastically lower the overall cost.
This result further reinforces the relevance of improving the overall system efficiency in the presence of higher reencoding errors.
Reducing the intrinsic cavity scattering losses from a state-of-the-art value towards ideal conditions, the tree cluster state generation rate $\Gamma_{\mathrm{tcs}}$ related to the cost function by $C\sim 1/\Gamma_{\mathrm{tcs}}$ (see supplementary information) rises by over an order of magnitude (blue curve in \subfigref{fig:Repeater}{b}).
Even in case of an efficiently fiber-coupled photon source without Purcell enhancement, this observation remains almost unchanged (red curve in \subfigref{fig:Repeater}{b}).
In presence of Purcell enhancement and intrinsic scattering losses of $2Q\approx\num{156000}$ corresponding to $\eta_\mathrm{SnV}=\num{94.7}\,\%$, the state generation rate rises from $\SI{1.1}{\mega\hertz}$ to $\SI{12.5}{\mega\hertz}$.
Though, already with an intrinsic state-of-the-art quality factor of $2Q\approx\num{17000}$ \citep{Mouradian2017}, the state generation rate reaches the $\si{\mega\hertz}$ regime.
Expected improvements in diamond nanofabrication in combination with our improved interface design thus put the optical requirements of the one-way repeater approach by J. Borregaard \textit{et al.} \citep{Borregaard2020} based on diamond vacancy centers embedded in Sawfish cavities within reach.

Although, the Sawfish cavity has been optimized for a SnV$^-$ center in diamond to demonstrate its properties and performance, we emphasize that the cavity design is adaptable to other diamond defect centers as well as to other material platforms hosting a quantum emitter. Scaling the cavity's dimensions adjusts its resonance frequency \citep{Joannopoulos2008}. Assuming the same cavity-to-waveguide and waveguide-to-fiber coupling efficiencies but the respective Debye-Waller factors, we present the expected system efficiencies for different diamond defect centers in \autoref{tab:Efficiencies}. Clearly, the Sawfish cavity raises the bulk Debye-Waller factor $DW^0$ of every emitter and thus provides high system efficiencies not only for the SnV$^-$. Particularly, the NV$^-$ being susceptible to surface charge traps \citep{Chakravarthi2021a, OrphalKobin2022} profits from an enlarged distance to surfaces as granted by the Sawfish design.

Having demonstrated advantages of the Sawfish cavity in simulations, its performance after diamond nanofabrication and as a building block for photonic integrated circuits demands further investigation. Concerning fabrication tolerances, predictions of the trained surrogate model necessitate verification by analyzing batches of fabricated cavities. To this point, it is not known to what extent rough surfaces possibly caused by diamond etching affect the Sawfish cavity's performance.

Optical losses in any spin-photon interface lead to a tremendously reduced efficiency in quantum information processing applications. This especially holds if an emitter has to repeatedly mediate entanglement for resource states involving hundreds of photons \citep{Borregaard2020} or if quantum memories become involved \citep{Bhaskar2020}. Whereas there are most efficient photon detectors with $\num{99.5}\,\%$ detection efficiency nowadays \citep{Chang_2021}, equally efficient spin-photon interfaces are still missing. Our system enables a spin-photon interface exceeding the efficiency of $\num{83}\,\%$ as needed by the PEPSI scheme \citep{Chen2021}, even with intrinsic state-of-the-art scattering losses. Consequently, we deem the Sawfish cavity design to be critical in interfacing quantum emitters and memories with near-unity efficiencies as required for scalable quantum networks.

\begin{table}[h]
    \small
    \setlength{\tabcolsep}{5pt}
	\centering
	\caption{Overall emitter-to-fiber efficiencies for different defect centers in diamond assuming the same waveguide $\left(\beta_\mathrm{WG}\right)$ and fiber $\left(\beta_\mathrm{F}\right)$ coupling efficiencies as found by simulations for the SnV$^-$ and either ideal ($F_\mathrm{P}=\num{270.2}$) or realistic state-of-the-art ($F_\mathrm{P}=\num{46.4}$) conditions. $\tau^0$ denotes the considered excited state lifetime in bulk diamond. $\tau^\mathrm{ideal}$ and $\tau^\mathrm{real}$ denote the calculated Purcell-enhanced lifetimes for ideal and realistic conditions, respectively.}
	\begin{tabular}{c|c|c|c|c}
		& NV$^-$ & SiV$^-$ & GeV$^-$ & SnV$^-$ \\
		\hline
		$\tau^0$ $[\si{\nano\second}]$ & $\num{12.2}$ \textnormal{\citep{Tamarat2006}} & $\num{1.7}$ \textnormal{\citep{Rogers2014a}} & $\num{3.8}$ \textnormal{\citep{Siyushev2017}} & $\num{4.5}$ \textnormal{\citep{Trusheim2020}} \\
		\hline
		$\tau^\mathrm{ideal}$ $[\si{\nano\second}]$ & $\num{1.3}$ & $\num{.01}$ & $\num{.03}$ & $\num{.04}$ \\
		\hline
		$\tau^\mathrm{real}$ $[\si{\nano\second}]$ & $\num{5.1}$ & $\num{.07}$ & $\num{.19}$ & $\num{.23}$ \\
		\hline
		$DW^0$ $[\%]$ & $\num{3}$ \textnormal{\citep{Santori2010}} & $\num{80}$ \textnormal{\citep{Neu2011}} & $\num{60}$ \textnormal{\citep{Palyanov2015}} & $\num{60}$ \textnormal{\citep{Goerlitz2020}} \\
		\hline
		$DW^\mathrm{ideal}$ $[\%]$ & $\num{89.3}$ & $\num{99.9}$ & $\num{99.6}$ & $\num{99.6}$ \\
		\hline
		$DW^\mathrm{real}$ $[\%]$ & $\num{59.4}$ & $\num{99.2}$ & $\num{98.0}$ & $\num{98.0}$ \\
		\hline
		$\eta^\mathrm{ideal}$ $[\%]$ & $\num{87.4}$ & $\num{97.6}$ & $\num{97.4}$ & $\num{97.4}$ \\ 
		\hline
		$\eta^\mathrm{real}$ $[\%]$ & $\num{53.8}$ & $\num{89.7}$ & $\num{88.6}$ & $\num{88.6}$ \\ 
	\end{tabular}
	\label{tab:Efficiencies}
\end{table}

\section{Methods}
As promising optically-active solid-state spin system, we exemplarily focus on negatively-charged tin vacancy centers in diamond (SnV$^-$). Their ZPL emission at $\SI{484.3}{\tera\hertz}$ \citep{Iwasaki2017} is coupled to the Sawfish cavities. Debye-Waller factors are calculated according to a formalism outlined in \citep{Li2015}. We investigate the system performing full 3d finite element (FEM) simulations with the software package JCMsuite \citep{JCMsuite}.

Firstly, the geometry parameters of the Sawfish cavity's central unit cell, like its lattice constant $a_0$, its thickness $2T$, its amplitude $2A_0$, and its gap width $2g$ are chosen to establish a large enough photonic band gap centered around the SnV$^-$ ZPL. Based on eigenmode computations, the higher-order lattice constants $a_i$ are then optimized for large quality factors using a Bayesian optimizer \citep{Schn:2019Benchmark} implemented in JCMsuite. The same optimizer is employed for designing the cavity-waveguide and waveguide-fiber interfaces. Investigating electromagnetic energy fluxes originating from a single dipole emitter placed at the cavity's center allows to analyze the system's efficiency.
Furthermore, we account for distributions of cavity design parameters by performing Monte Carlo sampling on a surrogate model composed of Gaussian processes and trained by means of machine learning. 
With the surrogate model, we gain insights into the cavity's robustness under displaced dipoles and fabrication tolerances. Finally, we apply the optimized system efficiencies to benchmark the potential performance of a one-way quantum repeater regarding the cost $C$, the required tree-cluster state size $N_\mathrm{ph}$, and the tree cluster state generation rate $\Gamma_{\mathrm{tcs}}$ following procedures detailed in \citep{Borregaard2020}. For details, see supplementary information.

\paragraph*{Data availability}
The datasets generated during the current study are available from the corresponding author on reasonable request.

\paragraph*{Code availability}
JCMsuite, Python, and MATLAB codes to perform and analyze the FEM and quantum repeater simulations as well as Python codes for Debye-Waller factor calculations are available from the corresponding author on reasonable request.

\paragraph*{Author contributions}
J.M.B. designed the Sawfish cavity and its waveguide coupling. M.P. developed the procedure to analyze the effect of fabrication tolerances. T.T. simulated the waveguide-to-fiber coupling. G.P. applied the Sawfish cavity to quantum repeater protocols. T.S. conceived the project. S.B. and T.S. supervised it. All authors contributed to the manuscript.

\paragraph*{Acknowledgement}
We thank Tommaso Pregnolato for fruitful discussions about realistic state-of-the-art fabrication scenarios as well as Kurt Busch, Martin Hammerschmidt, Lin Zschiedrich, and Philipp-Immanuel Schneider for helpful discussions about simulation methods.

\paragraph*{Funding}
This project is funded by the German Federal Ministry of Education and Research (BMBF) within the `DiNOQuant' project (No. 13N14921), the `QPIS' project (No. 16KISQ032K), and the `siMLopt' project (No. 05M20ZAA) as well as by the European Research Council within the ERC Starting Grant `QUREP' (No. 851810) and by the `EMPIR' program (No. 20FUN05 SEQUME) which is co-financed by the participating states and the European Union's Horizon 2020 research and innovation program. We acknowledge further funding from the Einstein Foundation Berlin within the frame of the `Einstein Research Unit on Quantum Devices'.

\paragraph*{Competing interests}
The authors declare no conflicting interest.

\printbibliography
\end{document}


\hyphenation{neg-a-tive-ly-charged wave-guide Borregaard nano-beam}

\title{`Sawfish' Photonic Crystal Cavity for Near-Unity Emitter-to-Fiber Interfacing in Quantum Network Applications}

\author[1,2]{Julian~M.~Bopp}
\author[3]{Matthias~Plock}
\author[1]{Tim~Turan}
\author[1]{Gregor~Pieplow}
\author[3,4]{Sven~Burger}
\author[1,2,*]{Tim~Schröder} 
\affil[1]{Humboldt-Universität zu Berlin, Department of Physics, 12489 Berlin, Germany}
\affil[2]{Ferdinand-Braun-Institut gGmbH, Leibniz-Institut für Höchstfrequenztechnik, 12489 Berlin, Germany}
\affil[3]{Zuse Institute Berlin (ZIB), 14195 Berlin, Germany}
\affil[4]{JCMwave GmbH, 14050 Berlin, Germany}
\affil[*]{Corresponding author: Tim Schröder, tim.schroeder@physik.hu-berlin.de}
\subtitle{Supplementary Information}
\maketitle

\section{Debye-Waller factor estimation}
\label{supp:DWfactor}
The procedure outlined in the following yields the Debye-Waller factors used in the main article to determine the cavity-enhanced ratio of light emitted into the respective emitter's zero-phonon line (ZPL).

\subsection{Group-IV vacancy centers}
The total decay rate from level $|3\rangle$ (\autoref{fig:SnVRateModel}) for a negatively-charged group-IV vacancy center in diamond (G4V) \citep{Bradac2019} is
\begin{equation}
	\gamma = \gamma_{32} + \gamma_{31} + \gamma_\mathrm{PSB}\,,
\end{equation}
where $\gamma_\mathrm{PSB} = \sum_k \gamma_{31_k}$ is the relaxation rate into the phonon sideband (PSB). 
In general, it is
\begin{align}
& \gamma_\mathrm{ZPL} = \gamma_{32} + \gamma_{31} = DW_\mathrm{G4V}^0 / \tau \,, \\
& \gamma_\mathrm{PSB} = (1- DW_\mathrm{G4V}^0) / \tau \,,
\end{align}
where $\tau = 1/\gamma$ is the excited state lifetime.  
The Debye-Waller factor in absence of Purcell enhancement is defined as 
\begin{equation}
	DW_\mathrm{G4V}^0 =\frac{\gamma_\mathrm{ZPL}}{\gamma} = \frac{\gamma_{31} + \gamma_{32}}{\gamma}\,.
\end{equation}
Tab.~1 of the main article lists $DW_\mathrm{G4V}^0$ for different G4Vs. It is assumed that $Q>\num{10000}$ at the wavelength of the $|3\rangle \rightarrow |1\rangle$ transition. This implies a cavity linewidth $\Delta\omega < \SI{50}{\giga\hertz}$, which is smaller than the ground state energy splittings of the emitters analyzed in this work. Thus, the only transition rate affected by Purcell enhancement with a Purcell factor $F_\mathrm{P}$ is
\begin{equation}
	\gamma_{31} \rightarrow \gamma_{31}(1 + F_\mathrm{P})\,.
\end{equation}
The Debye-Waller factor in presence of Purcell enhancement then becomes
\begin{align}
\begin{aligned}
	DW_\mathrm{G4V}&\left(F_\mathrm{P}\right) = \\ &\frac{\gamma_{31}F_\mathrm{P}\left(1-DW_\mathrm{G4V}^0\right) + DW_\mathrm{G4V}^0 \gamma_\mathrm{PSB}}{\gamma_{31}F_\mathrm{P}\left(1-DW_\mathrm{G4V}^0\right) + \gamma_\mathrm{PSB}}\,.
\end{aligned}
\end{align}
We estimate $\gamma_{31}$ and $\gamma_{23}$ from the reported G4V lifetimes (see Tab.~1 of the main article) by using the electronic structure model described e.g. in \citep{Trusheim2020} and the radiative rate of spontaneous emission given by
\begin{equation}
	\gamma_{ij} = \frac{\omega^3 n |\mu_{ij}|^2}{3 \pi \epsilon_0 \hbar c^3} \,.
\end{equation}
Here, $\omega$ is the transition frequency, $n$ the index of refraction and $|\mu_{ij}| = |\langle i| e \vec{r} |j \rangle|$ the magnitude of the transition dipole moment.

\fig[.4][Energy levels and spontaneous decay rates of a G4V$^-$ center in diamond.]{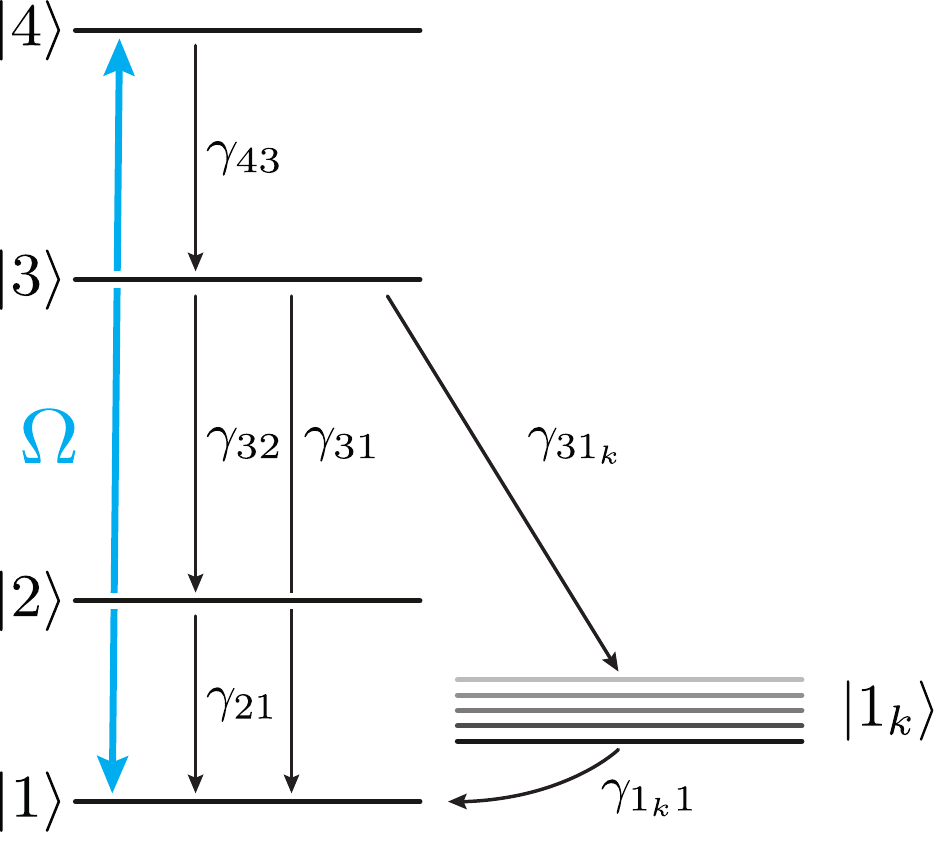}

\subsection{Nitrogen vacancy center}
The Debye-Waller factor for a negatively-charged nitrogen vacancy center in diamond (NV) \citep{Doherty2013} in presence of Purcell enhancement similarly becomes
\begin{equation}
DW_\mathrm{NV} = \frac{(1 + F_\mathrm{P})\gamma_\mathrm{ZPL}}{\gamma_\mathrm{PSB} + (1+F_\mathrm{P})\gamma_\mathrm{ZPL}} \,,
\end{equation}
where $\gamma_\mathrm{ZPL}$ is the rate of decay into the ZPL, $\gamma_\mathrm{PSB}$ the rate of decay into the PSB and $F_\mathrm{P}$ the Purcell factor. Both rates can be calculated using the excited state lifetime $\tau \approx \SI{12.2}{\nano\second}$ \citep{Tamarat2006} and the Debye-Waller factor of the unperturbed system
\begin{align}
& \gamma_\mathrm{ZPL} = DW^0/\tau \,, \\
& \gamma_\mathrm{PSB} = (1- DW^0)/\tau \,.
\end{align}

\section{Quantum repeater efficiency}
\label{supp:rep_eff}
We analyze the performance of the Sawfish cavity in the context of the quantum repeater scheme by J. Borregaard \textit{et al.} \citep{Borregaard2020}. The performance is assessed by optimizing the same cost function as defined in \citep{Borregaard2020} 
\begin{equation}
	C = \frac{1}{\Gamma_\mathrm{tcs} f p_\mathrm{trans}}\frac{m L_\mathrm{att}}{ \tau_\mathrm{ph} L} \,,
	\label{eq:costBorr}
\end{equation}
where $\Gamma_\mathrm{tcs}$ is the tree-cluster state generation rate, $f$ the secret-bit fraction of the transmitted qubits, $p_\mathrm{trans}$ the transmission probability, and $m$ the number of repeater stations. Applying exactly the same assumption made in \citep{Borregaard2020}, furthermore we define the optical fiber attenuation length $L_\mathrm{att} = \SI{20}{\kilo\meter}$, the photon emission time $\tau_\mathrm{ph} = \SI{10}{\nano\second}$, as well as the total communication distance $L = \SI{1000}{\kilo\meter}$.

According to \citep{Borregaard2020}, $1/C$ can be interpreted as the secret key rate in units of the photonic qubit emission time per repeater station and per attenuation length for a given total distance $L$.
We involve the same secret-bit fraction for distributing a secret key with a six-state variation of the BB84 protocol~\citep{Scarani_2009}
\begin{equation}
	f = 1 - g(Q) - Q - (1-Q) g\left(\frac{1- 3Q/2}{1- Q} \right) \,,
\end{equation}
where $Q = 2 \epsilon_\mathrm{trans} / 3$ is the qubit error rate and $g(x) = -x \log_2 x - (1-x) \log_2 (1 - x)$. Just as in \citep{Borregaard2020}, we approximate $\epsilon_\mathrm{trans} \approx (1 + m)\epsilon_\mathrm{r}$ with the error probability $\epsilon_\mathrm{r}$ of the reencoding step at the repeater station.
The transmission probability of a message qubit is
\begin{equation}
p_\mathrm{rans} = \eta_{e}^{m+1} \,,
\label{eq:trans_prob}
\end{equation}
where $\eta_\mathrm{e}$ is the transmission probability of an encoded qubit between repeater stations.
The recursive expression
\begin{align}
\eta_\mathrm{e} = [(1-\mu+\mu R_1)^{b_0} - (\mu R_1)^{b_0}](1 - \mu + \mu R_2)^{b_1}
\end{align}
with
\begin{equation}
R_k = 1 - [1 - (1-\mu)(1- \mu + \mu R_{k+2})^{b_{k+1}}]^{b_k}
\end{equation}
and $R_{d+1} = 0$, $b_{d+1}=0$, as well as $\mu = 1-\eta$ determines $\eta_\mathrm{e}$. The $b_k$ correspond to the tree cluster state with a branching vector $\vec{b} = [b_0,\dots, b_n]$.
In general, it is
\begin{equation}
\eta = \eta_\mathrm{detection}\eta_\mathrm{emitter} \times \exp\left(- L_0/L_\mathrm{att}\right) \,,
\end{equation}
where $L_0$ is the distance between adjacent repeater stations. 
The emitter-to-fiber collection efficiency
\begin{equation}
\eta_\mathrm{emitter} = \beta_\mathrm{C}\times\beta_\mathrm{WG}\times DW_\mathrm{emitter}\left(F_\mathrm{P}\right)\times\beta_\mathrm{F}
\end{equation}
is the crucial quantity in our analysis. $DW_\mathrm{emitter}$ is defined in \autoref{supp:DWfactor} Definitions of the coupling efficiencies $\beta$ can be found in the main article. The emitter-to-fiber collection efficiency is optimized by tuning the cavity design and the waveguide-to-fiber coupling geometry. We assume a detection efficiency of $\eta_\mathrm{detection} = \num{99.5}\,\%$ \citep{Chang_2021}.

The analysis performed in the main article is confined to tree cluster states with two levels $\vec{b}= [b_0, b_1, b_2]$.
The tree cluster generation rate $\Gamma_\mathrm{tcs}$ is given by 
\begin{align}
1/\Gamma_\mathrm{tcs} &= b_0[100 + b_1(1 + b_2)]\tau_\mathrm{ph} \nonumber\\ 
&+ b_0 [ 3 + b_1(1 + b_2)]\tau_\mathrm{CZ} \,,
\label{eq:supp-rate}
\end{align}
where $\tau_\mathrm{CZ} = \num{10} \tau_\mathrm{ph}$ is the controlled-Z gate time.  
The minimization of the cost function (equation \eqref{eq:costBorr}) is performed for a minimal distance of $L_\mathrm{min} = \SI{1}{\kilo\meter}$ per repeater station and maximally $\num{1000}$ photons in the tree-cluster state. The minimization returns the smallest cost $C_\mathrm{min}$ for a range of repeater stations and tree-cluster state configurations.

\section{FEM simulations}
We conduct full 3d finite element (FEM) simulations with the commercial finite element Maxwell solver software package JCMsuite \citep{JCMsuite}. The same software is used for the Bayesian optimization.

\subsection{Simulation types}
\label{sec:sim_type}
Either we calculate resonance modes of the investigated nanostructures or we apply a scattering approach where a dipole with a specific frequency and orientation is placed at a position within the 3d model.

Solving resonance problems typically yields the found eigenmodes' complex resonance frequencies $\tilde{\omega_i}$ besides their spatial electric field distribution. Quality factors $Q_i$ being defined as the ratio of a cavity's center frequency to its bandwidth thus become \citep{Kao2008}
\begin{align}
Q_i = \frac{1}{2}\left|\frac{\operatorname{Re}(\tilde{\omega_i})}{\operatorname{Im}(\tilde{\omega_i})}\right| \,.
\end{align}
Integrating the electric field intensity of the respective eigenmode leads to its mode volume $V_i$ after normalization \citep{Vuckovic2001}. Although the mode volume of a waveguide-coupled cavity is not precisely defined due to unknown integration bounds, we still consider this approach valid for two reasons. Firstly, by far the highest electric field intensities are reached in the cavity's center which renders the amount of light coupled to an attached waveguide almost negligible (which was confirmed by simulations). Secondly, since $F_\mathrm{P}\sim Q/V$ \citep{Purcell1946}, overestimated mode volumes would underestimate our Purcell factors turning them likely to be even higher in reality. Furthermore, resonance problems are used to determine the band structure of periodic structures consisting of \qt{Sawfish} unit cells.

In contrast, we use scattering approaches to investigate energy fluxes. These fluxes provide insights into the waveguide coupling efficiencies $\beta_\mathrm{WG}$ of asymmetric cavities and into waveguide-to-fiber coupling efficiencies $\beta_\mathrm{F}$. A dipole's radiation is propagated through the entire 3d model by the FEM solver. Inbuilt post-processes allow to calculate the dipole's totally emitted power as well as to integrate the Poynting vector across a specified surface within the 3d model. Integrating the Poynting vector across a waveguide or a fiber cross section takes fractions of the propagating optical mode into account which are guided in air. Comparing the dipole's totally emitted power with the integrated Poynting vector yields the transmission efficiency of a specified interface within the model. The Purcell factor can be alternatively estimated as the ratio of the dipole's totally emitted power to the power it emits in bulk. For the dipole displacement and fabrication uncertainty analysis, we utilize this strategy.

\subsection{Numerical uncertainties in finite element models}
\label{sec:fem_convergence}
Numerical simulations are generally associated with uncertainties that depend on the
discretization accuracy of the problem, i.e. for finite element simulations on the choice of the mesh
discretization size $h$ and the employed finite element polynomial degree $p$. For complex finite element models,
determining the magnitude of the numerical uncertainties arising from a specific choice of $h$ and $p$ is a challenging task. This is due to the fact that a full convergence analysis is often impossible owing to the large memory
requirements for simulations with small element sizes $h$ and high expansion
order $p$.

In order to perform a convergence analysis of the Sawfish cavity, we consider the
cavity with zero amplitude but increased gap size, i.e. $A_{0} = \num{0}$ and $g
\to g + A_{0}$. This approach effectively turns the cavity into a waveguide. The
properties of this translation-invariant structure can be obtained either by solving the
propagating mode problem of the waveguide's 2d cross section or the resonance mode problem of the invariant 3d
system~\citep{pomplun2007adaptive,VCSEL_convergence}.

For the propagating mode problem, we determine the effective refractive index
$n_{\mathrm{eff}}$, which depends on the geometry and the employed wavelength
$\lambda_{0} = \SI{618.83}{\nano\meter}$ of the unperturbed dipole. To estimate the relative uncertainty of $n_{\mathrm{eff}}$, we increase the finite
element polynomial degree and calculate the relative deviation from the most
accurate result $\epsilon_{\mathrm{rel,2d}}$. The relatively small number of
unknowns of the propagating mode problem allows to calculate the result with
high accuracy, thereby providing a reference solution.

The resonance mode problem is parameterized by the Bloch vector with
amplitude $|\vec{k}_{\mathrm{B}}| = 2\pi n_{\mathrm{eff}} / \lambda_{0}$, with
$n_{\mathrm{eff}}$ as found in the propagating mode problem. The fundamental
eigenmode $\omega_{\mathrm{fund}}$ of the resonance mode problem is found close
to $\omega_{\mathrm{guess}} = 2 \pi c_{0} / \lambda_{0}$. From
$\omega_{\mathrm{fund}} \approx \omega_{\mathrm{guess}}$, we determine
$\lambda_{\mathrm{fund}}$, and in turn the relative numerical
uncertainty of the invariant 3d system as $\epsilon_{\mathrm{rel,3d}} \approx
|\lambda_{\mathrm{fund}} - \lambda_{0} | / \lambda_{0}$.

\begin{figure}[h]
  \centering
  \includegraphics{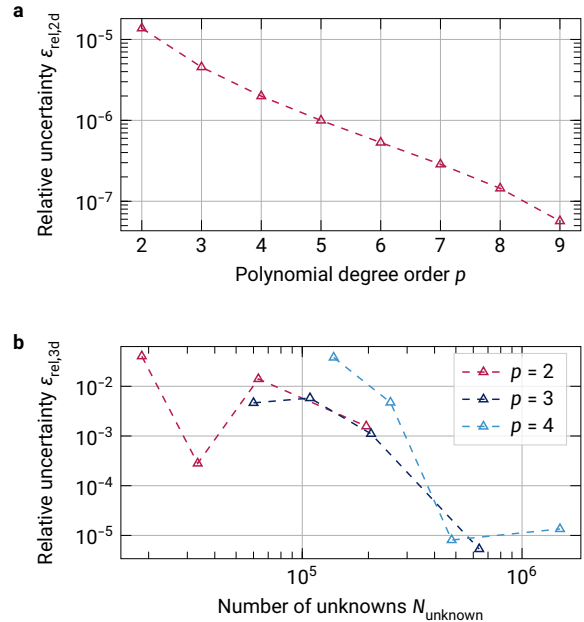}
  \caption{\textbf{Relative numerical uncertainties $\epsilon_{\mathrm{rel}}$ of the simplified finite element model.}
  \subfiglabel{a} depicts the uncertainty depending on the polynomial degree order $p$ for the 2d case and constant $h$.
  \subfiglabel{b} shows its dependence on different mesh discretizations $h$ for the 3d case and constant $p$. Smaller mesh discretizations $h$ lead to a larger number of unknowns $N_\mathsf{unknown}$ having to be determined. Dashed lines serve as guides to the eye.
   }
  \label{fig:fem_errors}
\end{figure}

The convergence analysis of the invariant 3d system is performed exploiting two mirror symmetry axes, thereby reducing the
number of unknowns by a factor of four. With $T =
\SI{133}{\nano\meter}$, $g = \SI{76}{\nano\meter}$ and a mesh discretization
size of $h = \SI{15}{\nano\meter}$, the fundamental mode of
the 2d system reveals a relative uncertainty of
$\epsilon_{\mathrm{rel,2d}} \approx \num{5e-8}$.
\autoref{fig:fem_errors} shows the results for calculating the same fundamental mode in the 2d as well as in the 3d system with a system length of $L = \num{10} \lambda_{0}$ for
different finite element polynomial degrees $p \in \{ 2, 3, 4 \}$ and mesh
discretizations $h \in \{ \lambda_0, \lambda_0 / 2, \lambda_0 / 4, \lambda_0 / 8
\}$. We observe that the uncertainty saturates at
approximately $\epsilon_{\mathrm{rel,3d}} \approx \num{1e-5}$, which can be
achieved by choosing $h = \lambda_0 / 8$ and $p = 3$.

\begin{figure}[h]
  \centering
  \includegraphics{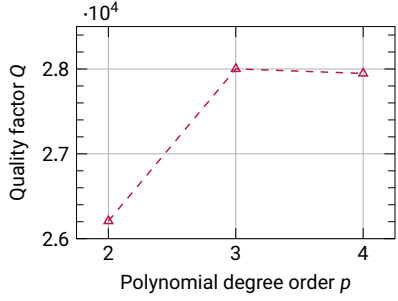}
  \caption{Dependence of a symmetric Sawfish cavity's quality factor $Q$ with $N=\num{12}$ on the polynomial degree order $p$ in a 3d FEM simulation.}
  \label{fig:fem_q}
\end{figure}

To verify the convergence results considering a real Sawfish cavity, we calculate the quality factor $Q$ for a symmetric cavity with $N=\num{12}$ unit cells at either side of the cavity's center and different finite element polynomial degrees $p$ (\autoref{fig:fem_q}). In agreement with the parameter choice targeting $\epsilon_{\mathrm{rel,3d}} \approx \num{1e-5}$, the quality factor stays constant for $p\geq\num{3}$.

\subsection{Cavity resonance estimation}
\label{sec:res_freq_unc}
Due to numerical uncertainties, the resonance frequencies of (symmetric and asymmetric) cavities deviate slightly from each other applying either the resonance problem or the scattering problem approach. We observe deviations up to approximately $\SI{10}{\giga\hertz}$ in between both approaches. In the worst case and for high quality factors and thus narrow linewidths, a resonance frequency determined by the resonance problem approach is not resonant if the scattering problem approach is applied with that frequency.

\begin{figure}[h]
  \centering
  \includegraphics{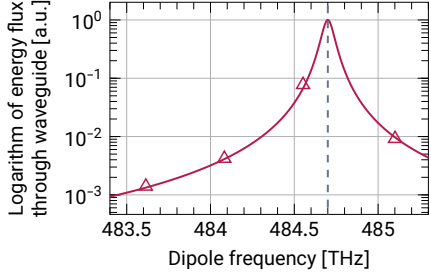}
  \caption{Exemplary Lorentzian fit of the frequency-dependent energy flux through the waveguide attached to a waveguide-coupled Sawfish cavity for estimating the \qt{scattering resonance frequency}.}
  \label{fig:res_freq_unc}
\end{figure}

To overcome this issue, we estimate the \qt{scattering resonance frequency} first for each scattering problem we are solving. At resonance, the electromagnetic energy flux through the waveguide of a waveguide-coupled cavity is highest. By solving the scattering problem for four dipole frequencies close to the expected resonance frequency, we obtain the frequency-dependent cavity transmission. Fitting the transmission with a Lorentzian function reveals the scattering resonance frequency (\autoref{fig:res_freq_unc}). This frequency is now used to recompute the respective scattering problem at resonance.

\subsection{Dipole orientation}
Diamond substrates cut along the crystallographic $\left[100\right]$ direction cause dipoles to enclose an angle of $\alpha=\num{54.7}^\circ$ with each cartesian axis ($\left[111\right]$ direction) \citep{Hepp2014, Goerlitz2020}. In turn, the dipoles do neither optimally overlap with the desired TM-like cavity mode nor with the modes of a rectangular waveguide attached to the cavity.

\begin{figure*}[h]
  \centering
  \includegraphics{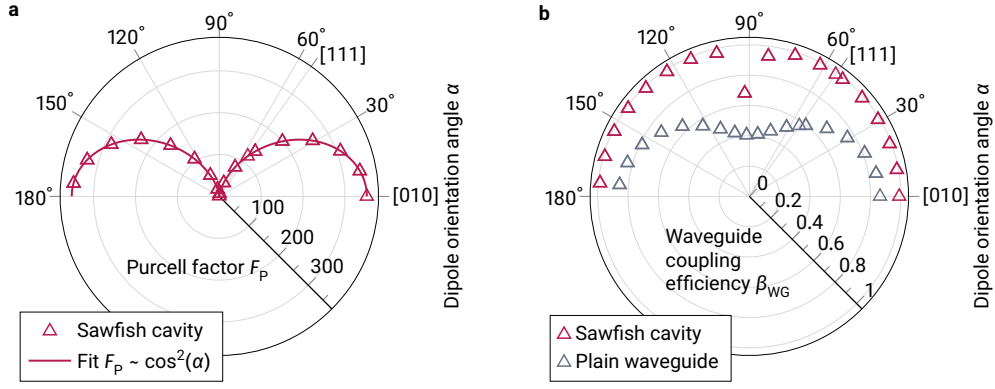}
  \caption{%
  Dependence of the Purcell factor $F_\mathrm{P}$ and the waveguide coupling efficiency $\beta_\mathrm{WG}$ on the rotation angle $\alpha$ of a dipole being rotated within a plane defined by the crystallographic [010] and [111] directions for a Sawfish cavity with parameters $N_\mathrm{s}=\num{23}$, $N_\mathrm{w}=\num{22}$, and $M=\num{19}$. The dipole possesses a fixed emission frequency of $\SI{484.47}{\tera\hertz}$. For symmetry reasons, data in the range $180^\circ <\alpha \leq 360^\circ$ is not plotted. For comparison, \subfiglabel{b} additionally shows the coupling efficiency of a dipole embedded into a plain, straight waveguide with the same rectangular cross section as the waveguide attached to the Sawfish cavity summing up the emission into both propagation directions (gray triangles).
  }
  \label{fig:wg_orientation}
\end{figure*}

For a dipole embedded into a Sawfish cavity with parameters $N_\mathrm{s}=\num{23}$, $N_\mathrm{w}=\num{22}$, and $M=\num{19}$, we examine the dipole orientation's influence by rotating the dipole within a plane spanned by the crystallographic $\left[111\right]$ ($\alpha=\num{54.7}^\circ$) and the $\left[010\right]$ ($\alpha=\num{0}^\circ$) direction (\autoref{fig:wg_orientation}). The main article describes respective observations on $F_\mathrm{P}\left(\alpha\right)$ and on $\beta_\mathrm{WG}\left(\alpha\right)$. If the dipole is, for comparison, embedded into a plain waveguide with the rectangular cross section of the waveguide attached to the tapered Sawfish cavity, two aspects change as depicted in \subfigref{fig:wg_orientation}{b}. Firstly, the maximally achievable coupling efficiency is about $\num{86}\,\%$ (summing up the emission into both propagation directions). This proves that a cavity not only enhances the emission of an embedded dipole. It is furthermore a tool to couple light emitted by a dipole most efficiently to a guided waveguide mode by accurately designing the tapered interface between the cavity and the waveguide. Secondly, a plain waveguide is more susceptible to non-ideally oriented dipoles. For a dipole oriented along the $\left[111\right]$ direction in a plain waveguide, the waveguide coupling efficiency is decreased to $\num{66.3}\,\%$ of its ideal value.

\subsection{Cavity-to-waveguide interface}
\begin{figure*}
  \centering
  \includegraphics[width=1.3\columnwidth]{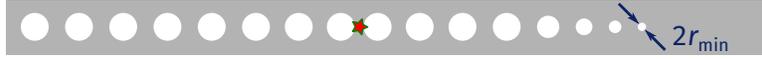}
  \caption{%
  Sketch of the established hole-based photonic crystal cavity design \citep{Mouradian2017, Evans2018, Rugar2021, Kuruma2021}. Arbitrarily small hole-like features cannot be fabricated. Particularly, diameters of hole-like features are limited to $2 r_\mathrm{min}$ within the cavity-to-waveguide interface.
  }
  \label{fig:tapered_hole_resonator}
\end{figure*}

To convert Bloch modes within the tapered cavity-to-waveguide interface into waveguide modes avoiding photon scattering, the cavity features have to taper off smoothly \citep{Palamaru2001, Sauvan2005}. The experimentally achievable minimal hole size is limited by fabrication constraints as sketched in \autoref{fig:tapered_hole_resonator}. To date, elliptical holes' minor axes of about $\SI{40}{\nano\meter}$ in about $\SI{200}{\nano\meter}$ thick diamond membranes have been reported \citep{Burek2017, Knall2022}. Thus, the arbitrary smooth conversion from Bloch to waveguide modes is not possible leading to increased scattering losses compared to the Sawfish design. Even if the corrugation features' tips in our Sawfish design suffered from such fabrication uncertainties, this would likely not affect the interface's performance since the electric field intensity is not localized at the tips (see cavity unit cells depicted in the main article).

\subsection{Waveguide-to-fiber interface}
The fiber-coupled Sawfish cavity also includes the well-established tapered waveguide-to-fiber interface \citep{Tiecke2015}, allowing us to quantify the total emitter-to-fiber interface efficiency $\eta$. To estimate the waveguide-to-fiber transmission $\beta_\mathrm{F}$, the waveguide's fundamental mode is launched from the untapered part of the waveguide into the tapered direction. An (arbitrary) power is assigned to it. After propagating and transitioning into a fundamental fiber mode via an evanescent waveguide-fiber supermode (see main article), the mode overlap between the electric field arriving at a fiber cross section and the fundamental fiber mode is computed as a power ratio.

\begin{figure*}[h]
  \centering
  \includegraphics{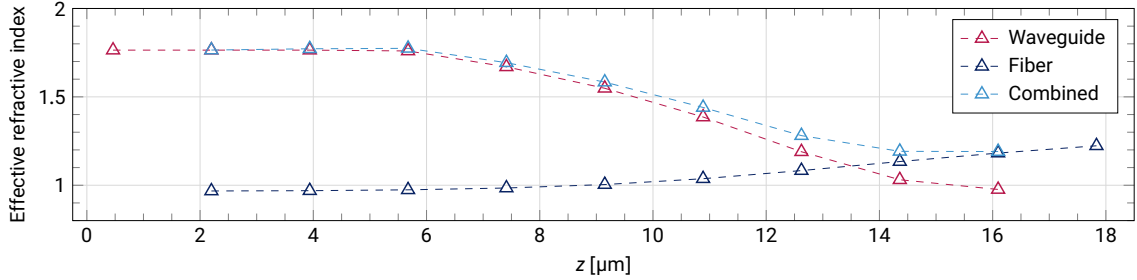}
  \caption{%
  Effective refractive indices within the frame of an adiabatic waveguide-to-fiber interface as attached to the Sawfish cavity. Modes guided in the tapered waveguide are displayed in red, modes guided in the tapered fiber in dark blue, and modes of the combined system in light blue. Dashed lines serve as guides to the eye.
  }
  \label{fig:fiber_index}
\end{figure*}

The effective refractive index of the transitioning optical mode is calculated at distinct cross sections of the 3d model employing propagating mode simulations performed with the software package JCMsuite \citep{JCMsuite}. \autoref{fig:fiber_index} shows effective refractive indices along the waveguide-to-fiber interface. They accurately resemble the situation presented in \citep{Tiecke2015} indicating an adiabatic transition.

\subsection{Bayesian optimization}
\label{sec:bayes_opt}
The optimized structures presented in the main manuscript are obtained by means of a
Bayesian optimization (BO)~\citep{movckus1975bayesian,mockus2012bayesian} method.
BO methods are sequential optimization methods that are very
efficient at optimizing black box functions or processes that are
expensive in terms of the consumed resources per
evaluation~\citep{jones1998efficient}. In BO methods, a stochastic surrogate
model, most often a Gaussian process (GP)~\citep{williams2006gaussian}, is
trained on observations drawn from an expensive black box function $f(\vec{p})$, where
$\vec{p} \in \mathcal{X} \subset \mathbb{R}^{N}$ and $f: \mathcal{X} \to
\mathbb{R}$. After training, the GP serves as a stochastic
predictor for the modeled function. In contrast to the modeled function itself,
GPs are usually much faster to evaluate. Additionally, when compared to
other machine learning methods such as deep neural networks~\citep{montavon2018methods,rasmussen2003gaussian}, the predictions made by a GP are usually easy to interpret~\citep{rasmussen2003gaussian}. This is due to the fact that the surrogate model's hyperparameters directly relate to
properties of the training data, like the mean, the variance, or the
length scales on which the data changes. The GP's predictions are used to iteratively
generate new sample candidates $\vec{p}$ for evaluating the expensive model function.
Sample candidates $\vec{p}$ are chosen to be effective for achieving the goal of
the optimization, e.g. for finding the global minimum of the modeled function.
The new values obtained from the expensive model retrain
the GP. This continues until the optimization budget is exhausted.

As noted, GPs are a key component in BO methods.
Being defined on the continuous domain $\mathcal{X}$, they extend
finite-dimensional multivariate normal distributions (MVNs) to an infinite
dimensional case~\citep{garnett_bayesoptbook_2022}. Where MVNs are specified by a mean vector
$\vec{\mu}$ and a positive (semi-)definite covariance matrix $\mat{\Sigma}$, GPs
are completely specified by a mean function $\mu: \mathcal{X} \to \mathbb{R}$
and a covariance kernel function $k: \mathcal{X} \times \mathcal{X} \to
\mathbb{R}$~\citep{williams2006gaussian}. We choose the commonly employed
constant mean function and the Mat\'ern
$\nicefrac{5}{2}$ kernel function~\citep{brochu2010tutorial}, i.e.
\begin{gather}
  \begin{aligned}[t]
    \mu(\vec{p}) &= \mu_{0} \,, \\
    k(\vec{p}, \vec{p}^{\prime}) &= \sigma_{0} \left( 1 + \sqrt{5}r +
      \frac{5}{3}r^2 \right) \times \exp{\left( -\sqrt{5}r \right)} \,,\\
  \end{aligned}
  \\
  \text{where} \quad r = \sqrt{\sum_{i=1}^{N} \left( \frac{ p_{i} -
        p^{\prime}_{i} }{l_{i}} \right)^{2}} \,.
\end{gather}
The hyperparameters $\mu_{0}$ and $\sigma_{0}$, as well as the length scales
$l_{1}, \dots, l_{N}$, are selected to maximize the likelihood of the observations
up to some maximum number of observations $W_{\mathrm{hyp}}$~\citep{garcia2018shape}.
Afterwards, only $\mu_{0}$ and $\sigma_{0}$ get updated. A GP that is trained on
$W$ function values $\vec{Y} = \left[ f(\vec{p}_{1}), \dots, f(\vec{p}_{W})
\right]\trans$ allows to make predictions in the form of a normal
distribution for each point $\vec{p}_{\ast}$ in the parameter space, i.e.
it predicts function values
\begin{gather*}
  \hat{f}(\vec{p}_{\ast}) \sim \mathcal{N}(\hat{y}(\vec{p}_{\ast}),
  \hat{\sigma}^{2}(\vec{p}_{\ast})) \,.
\end{gather*}
The values for the predicted mean $\hat{y}(\vec{p}_{\ast})$ and variance
$\hat{\sigma}^{2}(\vec{p}_{\ast})$ are defined as
\begin{align}
  \hat{y}(\vec{p}_\ast) &= \mu_0 + \vec{k}\trans (\vec{p}_\ast)
                               \mat{K}^{-1}[\vec{Y}-\mu_0 \vec{1}] \,,\\
  \hat{\sigma}^2(\vec{p}_\ast) &= \sigma_0^2 - \vec{k}\trans (\vec{p}_\ast)
                           \mat{K}^{-1} \vec{k}(\vec{p}_\ast) \,,
\end{align}
where $\vec{k}(\vec{p}_\ast) =
\left[k(\vec{p}_\ast,\vec{p}_1),\dots,k(\vec{p}_\ast,\vec{p}_{W})\right]\trans$
and $(\mat{K})_{i j} = k(\vec{p}_i,\vec{p}_j)$. For $\vec{p}_\ast$ very far from the
explored regions, the predictions approach the prior mean and variance
found during the hyperparameter optimization.

New sample candidates $\vec{p}_{W+1}$ are generated by selecting the parameter
that maximizes some utility function $\alpha(\vec{p})$.
Here, we consider the expected improvement (EI)
~\citep{jones1998efficient,mockus2012bayesian,garnett_bayesoptbook_2022} with
respect to the previously found smallest observed function value
$f_{\mathrm{min}}~=~\min \{ f(\vec{p}_{1}), \dots, f(\vec{p}_{W})\}$, i.e.
\begin{gather}
  \vec{p}_{W+1} = \underset{\vec{p} \in \mathcal{X}}{\arg\max}\, \alpha_{\mathrm{EI}}(\vec{p}) \,, \\
  \text{where} \quad \alpha_{\mathrm{EI}}(\vec{p}) = \ex{\min\left( 0,
      f_{\mathrm{min}} - \hat{f}(\vec{p}) \right)} \,.
\end{gather}

To perform the optimizations, we apply the analysis and optimization toolkit of
the commercial finite element Maxwell solver JCMsuite
\citep{Schn:2019Benchmark,Schn:2019GPR}.

\subsection{Impact of the fabrication tolerances and dipole displacements}
\label{sec:uncertainty_analysis}
By numerically optimizing the finite element models of the Sawfish cavity, we have
obtained a set of parameters $\vec{p}_\mathrm{opt}$ that theoretically leads to peak performance,
i.e. highest quality factors $Q$ or highest waveguide coupling efficiencies $\beta_\mathrm{WG}$.
Imperfect manufacturing processes, though, give rise to deviations from the targeted ideal
parameters. Thus, the actually realized parameters $\vec{p}_\mathrm{realized}$ scatter around $\vec{p}_\mathrm{opt}$
according to some probability distribution function (PDF) $\mathcal{F}$. This
naturally has a negative impact on the expected performance of the manufactured
device.

If the PDF of the manufacturing process $\mathcal{F}$ is known, the expected
performance reduction can be quantified by means of Monte Carlo sampling
techniques. In a naive approach, many samples from $\mathcal{F}$
are drawn and used to evaluate the finite element model $f(\vec{p})$. A
statistical analysis of the finite element results provides insight into the
expected performance reduction. However, the cost of a single evaluation of the full finite element
model renders this approach infeasible.

Instead of using the actual finite element model function to perform the Monte
Carlo sampling, we employ a surrogate model trained by machine learning based on Gaussian
processes~\citep{williams2006gaussian}, as introduced in \autoref{sec:bayes_opt}. Training is conducted on a comparatively small
set of model parameters $\vec{P}_{\mathrm{train}}$ and associated finite element
function values $f(\vec{P}_{\mathrm{train}})$. After training, the surrogate model serves as a
cheap-to-evaluate interpolator $\hat{f}(\vec{p})$ for the actual finite element
model. In \citep{rasmussen2003gaussian} and \citep{Plock2022}, a GP surrogate model similarly replaces an expensive model function during Monte Carlo sampling.

\subsubsection{Methodology}
\label{sec:mc_method}
We assume that the parameters $\vec{p}_\mathrm{realized}$ realized by the manufacturing 
process are distributed according to a multivariate
normal distribution (MVN) $\mathcal{F} = \mathcal{N}\left(
  \vec{\mu}_\mathrm{device}, \mat{\Sigma}_{\mathrm{device}} \right)$, with mean
$\vec{\mu}_{\mathrm{device}}$ and diagonal covariance matrix
$\mat{\Sigma}_{\mathrm{device}} = \diag\left( \sigma_1^2,\dots,\sigma_N^2
\right)$, i.e. $\vec{p}_\mathrm{realized} \sim \mathcal{F}$.
The set $\left\lbrace\sigma_{i}^2 \right\rbrace$ describes the variances of the individual
manufacturing process parameters. By assuming diagonality, we imply that the
outcome parameters are uncorrelated, i.e. that a parameter $p_i$ is not
dependent on another parameter $p_j$.

The GP surrogate model is trained on a set of $W_\mathrm{train}$ training parameters
$\vec{P}_{\mathrm{train}} = \{\vec{p}_1,\dots,\vec{p}_{W_\mathrm{train}} \}$,
that are used to
evaluate the expensive finite element model $f(\vec{p})$ and to generate the
training data
$\vec{Y} = \{f(\vec{p}_1),\dots,f(\vec{p}_{W_\mathrm{train}}) \}$. The training parameters
are drawn from a training distribution $\mathcal{G} = \mathcal{N}\left(
  \vec{\mu}_{\mathrm{train}}, \mat{\Sigma}_{\mathrm{train}} \right)$, with
$\vec{\mu}_{\mathrm{train}} = \vec{\mu}_{\mathrm{device}}$ and
$\mat{\Sigma}_{\mathrm{train}} = \kappa \times \mat{\Sigma}_{\mathrm{device}}$,
where $\kappa > 1$. Employing a MVN as a training distribution is advantageous since it favors
the more regularly sampled locations close to the mean
value of the manufacturing process distribution $\mathcal{F}$. As such, the
region around $\vec{\mu}_{\mathrm{device}}$ is trained more intensely which promotes a small
variance predicted by the GP surrogate. The available computational budget limits
the total number of training samples $W_\mathrm{train}$.
Mean values and variances predicted by the trained GP (c.f.
\autoref{sec:bayes_opt}) are then incorporated into the subsequent uncertainty impact
analysis.

In a Monte Carlo sampling approach~\citep{robert1999monte}, we incrementally draw
a large number of $N_{\mathrm{tot}}$ samples
$\vec{P}_{\mathrm{sample}} = \{
\vec{p}_{1},\dots,\vec{p}_{N_{\mathrm{tot}}} \}$ from the sampling distribution
$\mathcal{H} = \mathcal{F} = \mathcal{N}\left( \vec{\mu}_{\mathrm{device}},
  \mat{\Sigma}_{\mathrm{device}} \right)$. For these sample parameters,
the GP surrogate model is evaluated and the
function values $\hat{y}(\vec{p})$ predicted for the finite element model as
well as the associated variances $\hat{\sigma}^{2}(\vec{p})$ are calculated.
Being dependent on
random sample parameters, the predicted function values $\hat{\vec{Y}}_\mathrm{tot} =
\{ \hat{y}(\vec{p}_1), \dots, \hat{y}(\vec{p}_{N_{\mathrm{tot}}}) \} \sim
\mathcal{Q}$ are also random numbers that follow a statistical distribution
$\mathcal{Q}$. The same holds for the predicted variance samples
$\hat{\vec{S}}_\mathrm{tot} = \{
\hat{\sigma}^2(\vec{p}_{1}),\dots,\hat{\sigma}^2(\vec{p}_{N_{\mathrm{tot}}}) \}
\sim \mathcal{R}$. Analyzing certain percentiles of the distribution
$\mathcal{Q}$
allows to accurately quantify the impact of uncertainties in the
manufacturing process. By further analyzing the distribution of the predicted
variances $\mathcal{R}$,
we infer an estimate of the uncertainty introduced by using a surrogate model instead
of the expensive finite element model function.

For a Gaussian distribution, the $\num{50}$'th percentile describes the
mean which equals the median. The $\num{16}$'th and $\num{84}$'th percentiles are tied
to the lower and upper standard deviation, respectively. Accordingly, we investigate these
percentiles of $\mathcal{Q}$ by analyzing the samples $\hat{\vec{Y}}_\mathrm{tot}$. The median
$P_{50}(\hat{\vec{Y}}_{\mathrm{tot}})$ is given as the value for which we
estimate a probability of $\SI{50}{\percent}$ that the manufactured cavity will
have at least this value. Lower and upper standard deviations are calculated as
\begin{align}
  \sigma_{-}(\hat{\vec{Y}}_{\mathrm{tot}}) &= P_{50}(\hat{\vec{Y}}_{\mathrm{tot}}) -
                                             P_{16}(\hat{\vec{Y}}_{\mathrm{tot}}) \quad \text{and} \\
  \sigma_{+}(\hat{\vec{Y}}_{\mathrm{tot}}) &= P_{84}(\hat{\vec{Y}}_{\mathrm{tot}}) -
                                             P_{50}(\hat{\vec{Y}}_{\mathrm{tot}}) \,.
\end{align}
In order to quantify the uncertainty induced by using a surrogate model, we consider
the $\num{50}$'th percentile of $\mathcal{R}$, i.e. $\sigma_{\mathrm{GP}}^2 =
P_{50}(\hat{\vec{S}}_{\mathrm{tot}})$.
In contrast to ordinary Monte Carlo
sampling, where the expectation value and its variance are calculated directly,
our approach is capable of describing skewed distributions.

The error in Monte Carlo methods generally decreases as the number of samples increases.
To estimate this error, we consider the Monte Carlo error for ordinary Monte
Carlo sampling~\citep{murphy2012machine}
\begin{equation}
  \label{eq:mc_err}
  \sigma_{\mathrm{MC}}^2 \approx \frac{\var{\hat{\vec{Y}}_{\mathrm{tot}}}}{N_{\mathrm{tot}}} \,.
\end{equation}
It requires the variance and in turn also the expectation value of the
sample data $\hat{\vec{Y}}_\mathrm{tot}$, i.e.
\begin{gather}
  \label{eq:mc_var}
  \var{\hat{\vec{Y}}_{\mathrm{tot}}} = \frac{1}{N_{\mathrm{tot}}}
  \sum_{i=1}^{N_{\mathrm{tot}}}
  \left( \hat{y}(\vec{p}_i) - \ex{\hat{\vec{Y}}_{\mathrm{tot}}} \right)^2
  \\
  \text{and} \quad
  \label{eq:mc_ex}
  \ex{\hat{\vec{Y}}_{\mathrm{tot}}} = \frac{1}{N_{\mathrm{tot}}}
  \sum_{i=1}^{N_{\mathrm{tot}}}\hat{y}(\vec{p}_i)
  \,.
\end{gather}
While the Monte Carlo error can be reduced by sampling the
surrogate more often, the uncertainty introduced by the surrogate can only be reduced
by increasing the number of training points $W_\mathrm{train}$. The Monte Carlo error and the
surrogate uncertainty are combined into a compound uncertainty for the median
\begin{equation}
  \sigma_{\mathrm{Median}} = \sqrt{\sigma_{\mathrm{MC}}^{2} + \sigma_{\mathrm{GP}}^{2}} \,.
\end{equation}

The expected performance of the cavity is given in terms of these values, i.e.
\begin{equation}
  \overline{f} = \left( P_{50}(\hat{\vec{Y}}_{\mathrm{tot}}) \pm \sigma_{\mathrm{Median}} \right)^{\sigma_{+}(\hat{\vec{Y}}_{\mathrm{tot}})}_{\sigma_{-}(\hat{\vec{Y}}_{\mathrm{tot}})} \,.
\end{equation}
\autoref{alg:mc_int} outlines the complete procedure.

\begin{algorithm}[h]
  \caption{Determines the expected cavity performance for known
    manufacturing process uncertainties by means of Monte Carlo sampling of a
    trained Gaussian process surrogate model.}
  \label{alg:mc_int}
  \begin{algorithmic}
    \Procedure{MC simulation}{Predicted GP mean $\hat{y}(\vec{p})$, predicted GP
      variance $\hat{\sigma}^{2}(\vec{p})$, sample distribution $\mathcal{H}$}

    \State $\sigma_{\mathrm{rel,MC}} \gets \infty$
    \Comment \parbox[t]{.5\linewidth}{Relative MC error}

    \State $\sigma_{\mathrm{lb}} \gets \num{1e-3}$
    \Comment \parbox[t]{.5\linewidth}{Lower error bound}

    \State $N_{\mathrm{tot}} \gets \num{0}$
    \State $\Delta N \gets \num{1000}$
    \Comment \parbox[t]{.5\linewidth}{Sample increment}
    \State $N_{\mathrm{min}} \gets \num{50000}$
    \Comment \parbox[t]{.5\linewidth}{Minimum number of samples to be drawn}

    \State $\hat{\vec{Y}}_{\mathrm{tot}} \gets \emptyset$
    \State $\hat{\vec{S}}_{\mathrm{tot}} \gets \emptyset$
    
    \While{$( \sigma_{\mathrm{rel,MC}} \geq \sigma_{\mathrm{lb}}$ \textbf{and}
      $N_{\mathrm{tot}} < N_{\mathrm{min}} )$}

    \State $\vec{P}_{\mathrm{sample}} \gets$ draw $\Delta N$ samples from $\mathcal{H}$

    \State $\hat{\vec{Y}} \gets \hat{y}(\vec{P}_{\mathrm{sample}})$

    \State $\hat{\vec{S}} \gets \hat{\sigma}^{2}(\vec{P}_{\mathrm{sample}})$

    \State $\hat{\vec{Y}}_{\mathrm{tot}} \gets \hat{\vec{Y}}_{\mathrm{tot}} \cup \hat{\vec{Y}}$
    \State $\hat{\vec{S}}_{\mathrm{tot}} \gets \hat{\vec{S}}_{\mathrm{tot}} \cup \hat{\vec{S}}$
    \State $N_{\mathrm{tot}} \gets N_{\mathrm{tot}} + \Delta N$

    \State Calculate $\sigma_{\mathrm{MC}}(\vec{Y}_{\mathrm{tot}})$
    \Comment \parbox[t]{.3\linewidth}{\autoref{eq:mc_err}}
    
    \State $\sigma_{\mathrm{rel,MC}}(\vec{Y}_{\mathrm{tot}}) \gets
    \sigma_{\mathrm{MC}}(\vec{Y}_{\mathrm{tot}}) / P_{50}(\vec{Y}_{\mathrm{tot}})$

    \EndWhile

    \State $\sigma_{\mathrm{GP}}^{2} \gets P_{50}(\hat{\vec{S}}_{\mathrm{tot}})$

    \State $\sigma_{\mathrm{Median}} \gets \sqrt{\sigma_{\mathrm{GP}}^{2} +
      \sigma_{\mathrm{MC}}^{2}}$

    \State $\sigma_{+}(\vec{Y}_{\mathrm{tot}}) \gets
    P_{84}(\vec{Y}_{\mathrm{tot}}) - P_{50}(\vec{Y}_{\mathrm{tot}})$

    \State $\sigma_{-}(\vec{Y}_{\mathrm{tot}}) \gets
    P_{50}(\vec{Y}_{\mathrm{tot}}) - P_{16}(\vec{Y}_{\mathrm{tot}})$

    \State \textbf{return} $\left( P_{50}(\hat{\vec{Y}}_{\mathrm{tot}}) \pm
      \sigma_{\mathrm{Median}}
    \right)^{\sigma_{+}(\hat{\vec{Y}}_{\mathrm{tot}})}_{\sigma_{-}(\hat{\vec{Y}}_{\mathrm{tot}})}$
    
    \EndProcedure
  \end{algorithmic}
\end{algorithm}

\subsubsection{Impact of dipole displacements}
\label{sec:dipole_displacement}
The exact position of the dipole emitter within the cavity impacts
the expected Purcell factor $F_{\mathrm{P}}$ and the waveguide coupling efficiency
$\beta_{\mathrm{WG}}$. To quantify the position's influence, the method introduced in
\autoref{sec:mc_method} is applied. For the manufacturing process distribution, it
is assumed that the placement of the dipole can be achieved with a
standard deviation of $\SI{25}{\nano\meter}$ in each cartesian direction around
the cavity's center.
A training dataset with $\num{209}$ data points $\vec{P}_{\mathrm{train}} \sim
\mathcal{N}(\vec{\mu}_{\mathrm{train}}, \mat{\Sigma}_{\mathrm{train}})$ is
generated. Here, $\vec{\mu}_{\mathrm{train}} = \vec{0}$ and
$\mat{\Sigma}_{\mathrm{train}} = \diag( \SI{35}{\nano\meter},
\SI{35}{\nano\meter}, \SI{35}{\nano\meter})^{2}$, i.e.
relying on a standard deviation of $\SI{35}{\nano\meter}$ in each cartesian direction,
the training dataset encloses the assumed placement distribution.
The expensive finite element model function is evaluated using
$\vec{P}_{\mathrm{train}}$. Five training parameters in
$\vec{P}_{\mathrm{train}}$ are removed since their results deviate strongly from the local
average over adjacent samples. The GP surrogate now originates from
the $\num{204}$ remaining training parameters and results.

\begin{figure}[h]
  \centering
  \includegraphics{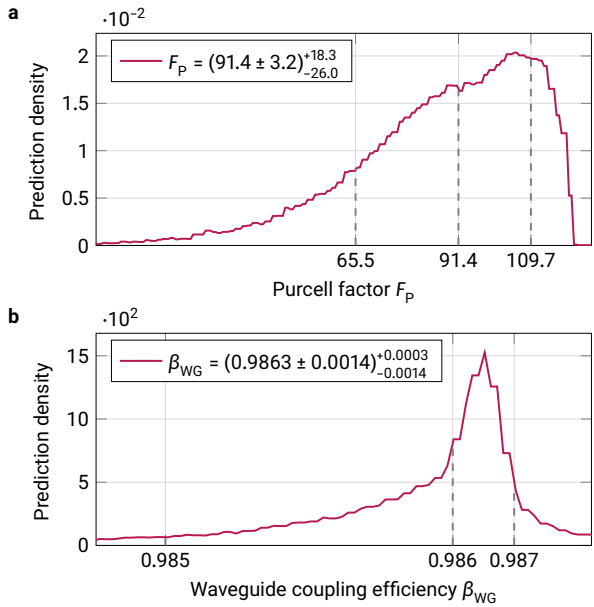}
  \caption{Expected cavity performance regarding the Purcell factor
    $F_\mathrm{P}$ \subfiglabel{a} and the waveguide coupling efficiency $\beta_\mathrm{WG}$
    \subfiglabel{b} for dipoles displaced according to a manufacturing process distribution
    $\mathcal{F}$. The values given on the abscissa relate to the respective
    median minus standard deviation, median, and median plus standard
    deviation.}
  \label{fig:dipole_displacement_monte_carlo}
\end{figure}

Accordingly, the sampling parameters $\vec{P}_{\mathrm{sample}}$
to evaluate the GP surrogates in the Monte Carlo sampling approach are
drawn from a MVN with $\vec{\mu}_{\mathrm{sample}} = \vec{0}$ and
$\mat{\Sigma}_{\mathrm{sample}} = \diag( \SI{25}{\nano\meter},
\SI{25}{\nano\meter}, \SI{25}{\nano\meter})^{2}$.
\autoref{fig:dipole_displacement_monte_carlo} depicts the
distributions of the predicted Purcell factor $F_{\mathrm{P}}$ and the waveguide
coupling efficiency $\beta_{\mathrm{WG}}$.

For the Purcell factor $F_{\mathrm{P}}$, we predict a value of
\begin{equation*}
  \overline{F}_{\mathrm{P}} = \left( \num{91.4} \pm \num{3.2} \right)^{+\num{18.3}}_{-\num{26.0}}
  \,,
\end{equation*}
where the standard deviation of the median $\sigma_{\mathrm{Median}}$ consists of
the surrogate uncertainty $\sigma_{\mathrm{GP}} = \num{3.2}$ and the numerical Monte
Carlo error $\sigma_{\mathrm{MC}} = \num{0.1}$. Likewise, we obtain $\beta_{\mathrm{WG}}$
\begin{equation*}
  \overline{\beta}_{\mathrm{WG}} = \left( \num{0.9863} \pm \num{0.0014} \right)^{+\num{0.0003}}_{-\num{0.0014}} 
  \,,
\end{equation*}
where the standard deviation of the median $\sigma_{\mathrm{Median}}$ consists of
the surrogate uncertainty $\sigma_{\mathrm{GP}} = \num{0.0014}$ and the numerical
Monte Carlo error $\sigma_{\mathrm{MC}} = \num{0.0001}$.

\subsubsection{Impact of fabrication tolerances}
\label{sec:fabrication_uncertainties}
Similarly, the exact geometry of the manufactured cavity
affects the expected Purcell factor
$F_\mathrm{P}$ and the waveguide coupling efficiency $\beta_{\mathrm{WG}}$.
For the manufacturing process distribution, we assume fabrication parameters
scattered around a desired mean value of $\vec{p}_{\mathrm{opt}} = \left[
  \Delta T,\allowbreak{} \delta,\allowbreak{} \Delta g \right] = \left[
  \SI{0}{\nano\meter},\allowbreak{} \SI{90}{\degree},\allowbreak{} \SI{0}{\nano\meter} \right]$
with uncertainties $\mat{\Sigma}_{\mathrm{sample}} = \diag(
\SI{0.8}{\nano\meter}, \SI{0.1}{\degree}, \SI{0.8}{\nano\meter})^{2}$.
A training dataset consisting of $\num{349}$ data points
$\vec{P}_{\mathrm{train}} \sim \mathcal{N}(\vec{\mu}_{\mathrm{train}},
\mat{\Sigma}_{\mathrm{train}})$ is produced. Here, $\vec{\mu}_{\mathrm{train}} =
\vec{p}_{\mathrm{opt}}$ and $\mat{\Sigma}_{\mathrm{train}} = \diag(
\SI{3.75}{\nano\meter}, \SI{1.35}{\degree}, \SI{3.75}{\nano\meter})^{2}$, i.e. we select
a training distribution which encloses the assumed manufacturing process distribution.
The expensive finite element model function is evaluated using
$\vec{P}_{\mathrm{train}}$ to train the GP surrogate models, excluding $\num{74}$ samples for which the resonance frequency could not be determined according to \autoref{sec:res_freq_unc}.

\begin{figure}[H]
  \centering
  \includegraphics{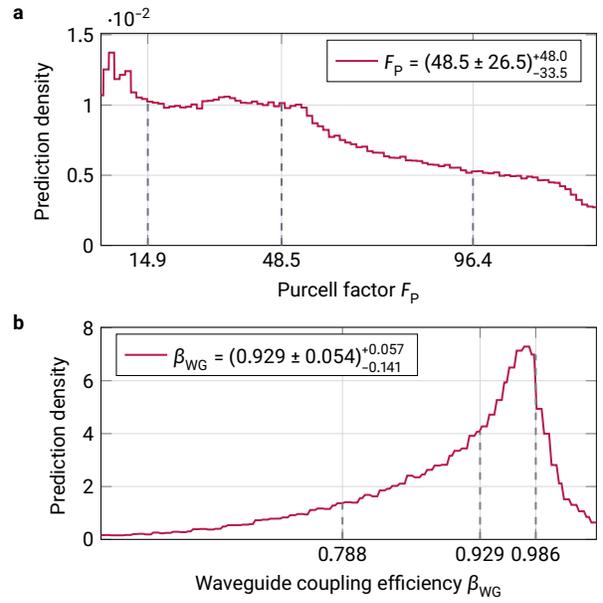}
  \caption{Expected cavity performance regarding the Purcell factor
    $F_\mathrm{P}$ \subfiglabel{a} and the waveguide coupling efficiency $\beta_\mathrm{WG}$
    \subfiglabel{b} for fabrication (geometry) parameters distributed according to a
    manufacturing process distribution $\mathcal{F}$. The values given on the
    abscissa relate to the respective median minus standard deviation, median,
    and median plus standard deviation.}
  \label{fig:fabrication_uncertainties_monte_carlo}
\end{figure}

The sampling parameters $\vec{P}_{\mathrm{sample}}$
to evaluate the GP surrogates in the Monte Carlo sampling approach are
drawn from a MVN with $\vec{\mu}_{\mathrm{sample}} =
\vec{p}_{\mathrm{opt}}$ and $\mat{\Sigma}_{\mathrm{sample}} = \diag(
\SI{0.8}{\nano\meter}, \SI{0.1}{\degree}, \SI{0.8}{\nano\meter})^{2}$.
Sampling parameters for which the surrogate evaluation leads to
$F_\mathrm{P}\leq\num{1}$ are excluded from the calculation of the predicted
values since they describe nonresonant (thus defective) cavities. The amount of samples discarded by this criterion is below $\num{5}\,\%$.
\autoref{fig:fabrication_uncertainties_monte_carlo} displays 
the distributions of the predicted Purcell factor $F_{\mathrm{P}}$
and the waveguide coupling efficiency $\beta_{\mathrm{WG}}$.

For the Purcell factor $F_{\mathrm{P}}$, this yields a predicted value of
\begin{equation*}
  \overline{F}_{\mathrm{P}} = \left( \num{48.5} \pm \num{26.5} \right)^{+\num{48.0}}_{-\num{33.5}}
  \,,
\end{equation*}
where the standard deviation of the median $\sigma_{\mathrm{Median}}$ consists of
the surrogate uncertainty $\sigma_{\mathrm{GP}} = \num{26.5}$ and the
numerical Monte Carlo error $\sigma_{\mathrm{MC}} = \num{0.1}$.
Likewise, it is
\begin{equation*}
  \overline{\beta}_{\mathrm{WG}} = \left( \num{0.929} \pm \num{0.054} \right)^{+\num{0.057}}_{-\num{0.141}} 
  \,,
\end{equation*}
where the standard deviation of the median $\sigma_{\mathrm{Median}}$ consists of
the surrogate uncertainty $\sigma_{\mathrm{GP}} = \num{0.054}$ and the numerical
Monte Carlo error $\sigma_{\mathrm{MC}} = \num{0.001}$.

\printbibliography